\documentclass[a4paper,11pt]{article}
\usepackage{jcappub} 
\usepackage{bigints}
\usepackage{fontawesome}
\usepackage{comment}
\usepackage[noabbrev]{cleveref}
\usepackage{braket}
\usepackage{xcolor}
\usepackage{bigints}
\usepackage{caption}

\definecolor{tiffany}{RGB}{79, 166, 158}

\title{Forward-modeled cosmo-astrophysical inference from binary black hole number counts with the Einstein Telescope}

\author[a,b]{Giovanni Antinozzi,}
\author[a,c]{Michele Bosi,}
\author[a,d,e]{Andrea Lapi,}
\author[f]{Simone Mastrogiovanni,}
\author[f,h]{Arianna Scarpa,}
\author[a,b,g]{and Mario Spera}

\affiliation[a]{SISSA, via Bonomea 365, I–34136 Trieste, Italy}
\affiliation[b]{National Institute for Nuclear Physics – INFN, Sezione di Trieste, I–34127 Trieste, Italy}
\affiliation[c]{Department of Physics, University of Trento, Via Sommarive 14, 38123 Povo (TN), Italy}
\affiliation[d]{Institute for Fundamental Physics of the Universe – IFPU, Via Beirut 2, I–34014 Trieste, Italy}
\affiliation[e]{Istituto di Radioastronomia – INAF/IRA, Via Piero Gobetti 101, I–40129 Bologna, Italy}
\affiliation[f]{INFN, Sezione di Roma, I-00185 Roma, Italy}
\affiliation[g]{Istituto Nazionale di Astrofisica – Osservatorio Astronomico di Roma, Via Frascati 33, I–00040, Monteporzio Catone, Italy}
\affiliation[h]{Dipartimento di Fisica, Sapienza Università di Roma, P.le A. Moro 5, Roma 00185, Italy}

\emailAdd{gantinoz@sissa.it, mspera@sissa.it}

\abstract{Gravitational-wave population studies are now a central tool for astrophysics and cosmology, but treat the merger-rate normalization as a nuisance parameter, as they primarily target the binary parameters. However, the merger rate carries additional information. We thus ask what binary black hole (BBH) number counts alone can constrain in the Einstein Telescope (ET) era. We forward-model the BBH merger rate by convolving a data-driven, metallicity-resolved cosmic star formation history with {\footnotesize\textup{\textrm{SEVN}}} population-synthesis simulations of isolated binaries, and run a mock inference on the Hubble constant $H_0$ and the common-envelope efficiency $\alpha_{\rm CE}$ for the 10\,km triangular ET. We predict about $10^5$ detections per year, roughly $60\%$ of all BBH mergers. Number counts alone cannot constrain $H_0$, but confine $(H_0,\alpha_{\rm CE})$ to a narrow degeneracy curve. If the model describes the true BBH population, $\alpha_{\rm CE}$ is recovered to about 5\% ($2\sigma$) for a fiducial $\alpha_{\rm CE}=3$ with a flat prior on $H_0$, or 40\% for $\alpha_{\rm CE}=1$; with $H_0$ known at the percent level, precision improves to better than 1\% and about 7\%. Calibration tests show well-calibrated posteriors for the correct model regardless of fiducial values, and reject misspecified models. Accuracy, however, is model dependent: changing the metallicity relation or delay-time distribution shifts the posterior by 25--100 times its Poisson width. Number counts thus offer an inexpensive consistency test of population models before a full spectral-siren analysis. In line with recent studies, our models overpredict the GWTC-5.0 local merger rate by a factor 4--9. The excess stems not from the star formation history but mainly from binaries undergoing common-envelope evolution after stable mass transfer, making these prescriptions a prime target for revision.}

\begin{document}
\maketitle
\flushbottom

\section{Introduction}\label{sec:intro}
Gravitational-wave astronomy has evolved rapidly over the past decade. A century after Einstein predicted gravitational waves (GWs) \cite{Einstein:1916_GW}, the first direct detection was made on September $14^{\rm th}$ 2015 \cite{Abbott:2016_firstGW}, opening a new observational window on the Universe. Since then, compact binary coalescences (CBCs) have enabled increasingly stringent tests of general relativity \cite{Abbott:2019GR_test1, Abbott:2021GR_test2, LVK:2021GR_test3, LVK:2026GWTC4_TGR1, LVK:2026GWTC4_TGR2, LVK:2026GWTC4_TGR3}, and have challenged our knowledge of stellar astrophysics \cite{Abbott:2020astro_chall_1, Abbott:2020astro_chall_2, Abbott:2020astro_chall_3, Abac:2025massive_GW, Abac:2024light_GW}. 

By the end of the second observing run of the LIGO-Virgo-KAGRA (LVK) collaboration, the catalog of GW detections was large enough to move beyond individual events and support the first population studies \cite{Abbott:2019pop_GWTC2,Abbott:2021GW_pop}. These studies connect compact-binary progenitors to binary-evolution models and test formation scenarios against observations \cite{Costa:2023BBH_rev, Tauris:2023binary_review}. Nevertheless, the formation channels and evolutionary pathways of binary black holes remain uncertain \cite{Spera:2022GW_binary_review, Callister:2024RevGWpop}. The growing catalog has revealed systems that challenge simple theoretical expectations, including events with component masses in regions traditionally associated with pair-instability processes or close to the proposed lower mass gap \cite{Abac:2025massive_GW, Abac:2024light_GW}. These systems highlight the substantial uncertainties that still affect theoretical predictions of compact binary formation and evolution. 

At the same time, GW observations provide a direct measurement of the luminosity distance, independently of the cosmic distance ladder, making them powerful standard sirens \cite{Schutz:1986standard_sirens}. Cosmological inference, however, requires both luminosity distance and redshift. GW signals directly encode the distance but not the redshift, because the measured masses are redshifted and the waveform alone cannot separate intrinsic mass from cosmological redshift \cite{Maggiore:2018gw_book}. Redshift information must therefore be obtained in other ways, typically through the identification of a host galaxy, which is challenging because GW events often have large sky-localization areas \cite{LVKAbac:2025GWTC4}. This has motivated two approaches: \emph{bright sirens}, which use an electromagnetic counterpart to identify the host galaxy and measure its redshift \cite{Abbott:2021H0_GW, Abbott:2017H0_BNS} and \emph{dark sirens}, which infer the redshift statistically or from the waveform, without a counterpart \cite{DelPozzo:2012Gal_cat_met, DelPozzo:2017BNS_tidal}. Within this class, \emph{spectral sirens} are a statistical dark-siren approach, in which features in the collective CBC distribution constrain astrophysical and cosmological parameters at once \cite{Taylor:2012Spectral_sirens1, Abbott:2023Spectral_siren_LVK}. 

The constraining power of spectral sirens depends critically on the population model, i.e. on how the binary black hole (BBH) merger rate is assumed to depend on masses, spins and redshift. This model encodes the astrophysical processes that shape binary formation and evolution and determines the predicted distribution of events. Three classes of population models are commonly used: \emph{parametric models}, which fit observationally motivated functional forms to the mass, spin and redshift distributions \cite{Fishbach:2017mass_spectrum_par}; \emph{non-parametric models}, which reconstruct these distributions directly from the data, e.g. with spline interpolants \cite{Mandel:2017mass_spectrum_spline} or stochastic processes \cite{Rinaldi:2022mass_spectrum_gausproc}; and \emph{population synthesis models}, which derive the compact-object population from stellar- and binary-evolution simulations. A key ingredient in these models is the cosmic star formation history, which sets both the normalization and the redshift evolution of the merger rate. This component has been modeled with parametric fits \cite{Madau:2017_SFRD}, cosmological hydrodynamical simulations \cite{Grudic:2023cosmosim_sfr}, and galaxy-based approaches \cite{Santoliquido:2022GalaxyRate, Artale2019, Artale2020a, Artale2020b}. 

Although both parametric and non-parametric methods provide flexible and powerful tools for population inference \cite{Fishbach:2017mass_spectrum_par, Talbot:2018par_mod_0, LVKAbac:2025GWTC4cosmo, LIGO:2025GWTC4, Gennari:2025par_mod_ev, Mandel:2017mass_spectrum_spline, Rinaldi:2022mass_spectrum_gausproc, Tiwari:2021nonpar_mod_0, Golomb:2023nonpar_mod_1, Edelman:2023nonpar_mod_2, Ray:2023non_par_mod_3, Heinzel:2024non_par_mod_4, Callister:2024non_par_mod_5, Pierra:2025nonpar_mod}, their connection to the underlying binary physics is indirect, and sufficiently flexible models may potentially produce features in the CBC spectrum that are difficult to interpret physically. In contrast, population-synthesis methods are currently more uncertain, but they are directly tied to our knowledge of stellar- and binary-evolution physics \cite{Stevenson:2022pop_synth_1, Zevin:2017pop_synth_2, Neijssel:2019pop_synth_3, Bouffanais:2019mix_f, Zevin:2021pop_synth_4}. In this work, we combine a population synthesis model with a data-driven description of the cosmic star formation history, based on empirical galaxy scaling relations, and we ask how much information BBH number counts alone can provide. 

By number counts we intend the total number of detected events, regardless of their individual parameters. The population likelihood combines a Poisson term for the number of detections $N_{\rm det}$ with the distribution of their parameters \cite{Mandel:2019hier_like2, Vitale:2022hier_like1, Thrane:2019intro_bayes_gw, Mastrogiovanni:2023icarogw} and the Poisson term is usually marginalized analytically with a scale-invariant prior $\pi(N_{\rm det}) \propto 1/N_{\rm det}$. This approach is justified as most of the information on the BBH mass and spin population distributions usually comes from values measured from individual binaries. However, in the context of astrophysical formation channels, we are also interested in checking if the overall physical rate predicted by the model is consistent with the observed one, thus requiring the information contained in the number counts.

The information content of number counts depends on the ability to probe the redshift evolution of the merger rate over a sufficiently wide range. Current detectors observe BBHs only at low redshift: fewer than $1\%$ of the BBHs in GWTC-5.0 lie above $z\approx1.4$ \cite{LVK:2026GWTC5_pop}. Their counts are therefore set by the local merger rate, a single number in which cosmology, star formation and binary evolution are all combined and cannot be separated. However, most mergers occur around and beyond the peak of the merger rate at $z\sim2$, which follows the peak of cosmic star formation \cite{Gray:2023zp}. This requires GW detectors that can observe binary black hole mergers out to $z \gtrsim 2$, placing these methods in the era of next-generation observatories such as the Einstein Telescope and Cosmic Explorer \cite{Abac:2025BlueBook,Branchesi:2023ET_design, Reitze:2019CE_1, Abbott:2017CE_2}. For this reason, we focus on one of the possible Einstein Telescope designs: the triangular 10 km configuration, which can probe the merger-rate evolution out to and beyond the peak of cosmic star formation \cite{Abac:2025BlueBook,  Branchesi:2023ET_design, Maggiore:2020ETD}.

We significantly refine the approach of \cite{Antinozzi:2024_number_paper}, adopting the same number-count likelihood, but substantially changing the model and the scope of the inference. First, the star formation history is no longer described by one of a set of analytic SFR models, but by a data-driven framework built from empirical galaxy scaling relations, which resolves the SFR density in host-galaxy stellar mass and metallicity. Second, the BBH merger rate is obtained by convolving this star formation history with {\footnotesize\textup{\textrm{SEVN}}} population-synthesis simulations of isolated binaries, so that its normalization and redshift evolution follow from explicit binary-evolution prescriptions. Third, the inference is no longer purely cosmological: we constrain jointly the Hubble constant and the common-envelope efficiency $\alpha_{\rm CE}$, targeting the degeneracy between them. Unlike approaches that infer phenomenological population parameters, we thus constrain a prescription of the simulations directly, through \emph{forward modeling}. This allows us to quantify how assumptions on binary evolution, metallicity and star formation propagate into the expected number of detections, which we finally test with dedicated calibration diagnostics.

The paper is organized as follows. In Subsec.~\ref{sec:SFRD}, we present our new framework for the star formation rate density and analyze the impact of varying cosmological parameters and fundamental metallicity relations. In Subsec.~\ref{sec:MRD}, we describe our simulated binary-star catalogs and our merger-rate-density model, focusing on the effect of different modeling assumptions, astrophysical parameters and fundamental metallicity relation (FMR) prescriptions. In Subsec.~\ref{sec:analysis}, we outline the hierarchical inference framework and specialize to the number-count analysis. In Sec.~\ref{sec:results} we present the number-count parameter estimation under different modeling assumptions and validate it on many mock datasets. In Sec.~\ref{sec:discussion}, we compare our models with the GWTC-5.0 local merger rate, and in Sec.~\ref{sec:summary} we summarize our results.

\section{Methods}
\subsection{The cosmic star formation rate density in a data-driven framework}\label{sec:SFRD}

In this section we construct a semi-empirical framework for the cosmic star formation rate (SFR) density and its evolution across cosmic time. We use star-forming galaxies as tracers of star formation, following the data-driven approach of \cite{Chruslinks:2019, Boco:2020pgp, Sgalletta:2025high_MRD}. Unlike \cite{Bosi:2025QG}, which followed only the progenitors of quiescent galaxies with a similar data-driven approach, we include all galaxies that are star forming at each epoch, as needed to model the full BBH population.

We assume a flat $\Lambda$CDM cosmology and let the Hubble constant $H_0$ vary. As fiducial cosmology (F) we adopt that of \cite{Weaver:2023SMF, Chruslinska:2021MS_SB, Popesso:2023MS_SFR}, $H_0=70$\,km\,s$^{-1}$\,Mpc$^{-1}$ and $\Omega_{\rm m,0} = 0.3$. We adopt $Z_{\odot} = 0.0153$ as solar metallicity \cite{Caffau:2011Z_sun}.

\subsubsection{Total SFR density}\label{subsec:SFRD_model}
We construct the cosmic SFR density, $\rho_{\rm SFR}(z)$, defined as the total star formation rate per unit comoving volume at redshift $z$, from two redshift-dependent ingredients. The first is the stellar mass function (SMF) of star-forming galaxies (SFGs), $\mathrm{d}N_{\rm SFG}/\mathrm{d}\log M_* \mathrm{d}V$, which describes the number density of star-forming galaxies per unit base 10 logarithmic stellar mass $\log M_*$ and per unit comoving volume $V_c$. The second ingredient is the conditional probability distribution $\mathrm{d}p/\mathrm{d}\log\psi(\log\psi|\log M_*,z)$, which gives the distribution of star formation rates $\psi$ for galaxies of stellar mass $M_*$ at redshift $z$. The SFR density is then obtained by integrating the galaxy number density weighted by the star formation rate and its distribution:
\begin{equation}
    \rho_{\rm SFR}(z) = \int \text{d}\log M_* \dfrac{\text{d} N_{\rm SFG}}{\text{d}\log M_*\text{d}V_c}(\log M_*, z)\int \text{d}\log \psi \dfrac{\text{d}p}{\text{d}\log\psi}(\log \psi|\log M_*, z)\,\psi\,.
    \label{eq:SFRD_tot}
\end{equation}
To compute the first term in the integral, we adopt the COSMOS2020 determination for the SMF by \cite{Weaver:2023SMF} as a double Schechter function:
\begin{equation*}
    \dfrac{\text{d} N_{\rm SFG}}{\text{d}\log M_*\text{d}V_c}(\log M_*, z) = \ln 10 e^{-10^{\log M_* - \log M^{\star}}}\Big[\phi_1\left(10^{\log M_* - \log M^{\star}}\right)^{\alpha_1 + 1} +
\end{equation*}
\begin{equation}
    + \phi_2\left(10^{\log M_* - \log M^{\star}}\right)^{\alpha_2 + 1}\Big]\,;
    \label{eq:SMF}
\end{equation}
where the parameters $(\log M^{\star}, \phi_{1,2}, \alpha_{1,2})$ are given in redshift bins in Tab. C.2 of \cite{Weaver:2023SMF}. Above $z=5.5$, where \cite{Weaver:2023SMF} no longer separate star-forming and quiescent galaxies, we extrapolate these parameters linearly in redshift. At these redshifts star-forming galaxies dominate the SMF, and \cite{Weaver:2023SMF} argue that extrapolating the star-forming relations is meaningful. The observed SMF is inferred assuming a fiducial cosmology, which enters through the definition of the comoving volume element. When varying cosmological parameters, the number of observed galaxies remains unchanged, but the associated comoving volume differs. To ensure consistency, the SMF must therefore be rescaled by the ratio of the comoving volume elements between the fiducial cosmology (F) and a generic cosmology (G):
\begin{equation}
    \dfrac{\mathrm{d}N_{\rm SFG}}{\mathrm{d}\log M_* \mathrm{d}V_{c,\rm G}} = \dfrac{\mathrm{d}N_{\rm SFG}}{\mathrm{d}\log M_* \mathrm{d}V_{c,\rm F}} \dfrac{H_{\rm G}(z)}{H_{\rm F}(z)}\left(\dfrac{r_{\rm F}(z)}{r_{\rm G}(z)}\right)^2\,,
    \label{eq:SFG_cosmo}
\end{equation}
with $H_\bullet(z)$ being the Hubble parameter and $r_\bullet(z)$ the comoving distance. 

We notice that, since the galaxy counts in a redshift shell do not depend on the galaxies intrinsic properties, a change in cosmology only rescales the comoving volume element in Eq.~(\ref{eq:SFG_cosmo}). This factor is independent of galaxy properties, so it applies identically to all the quantities derived from the SMF. For $H_0$ this is a redshift-independent rescaling, ${\rm SMF}_{\rm G}\propto H_0^3$, i.e. between $-37\%$ and $+49\%$ for $H_0\in[60,80]$\,km\,s$^{-1}$\,Mpc$^{-1}$. Varying $\Omega_{\rm m,0}$ in $[0.2,0.4]$ also changes the redshift evolution, through $H_{\rm G}(z)$ and $r_{\rm G}(z)$. The effect is negligible at low redshift and reaches 35--37\% at $z=10$. Overall, the SMF changes by less than 0.2\,dex over the explored ranges. Thus, cosmology is expected to affect the normalization and redshift evolution of the SFR density, but not its distribution over galaxy properties.

The conditional SFR distribution entering Eq.~(\ref{eq:SFRD_tot}) is modeled as a double log-normal. This choice is motivated by observations showing that, at fixed stellar mass and redshift, galaxies lie on a well-defined star-forming main sequence (MS), while a smaller fraction form stars at significantly enhanced rates and are classified as starburst (SB) systems \cite{Chruslinska:2021MS_SB}. We therefore describe the SFR distribution as the superposition of a main-sequence component and a starburst component, i.e.
 
\begin{equation}
    \dfrac{\text{d}p}{\text{d}\log \psi}(\log\psi|\log M_*,z) = \dfrac{\mathcal{N}_{\rm MS}}{\sqrt{2\pi \sigma_{\rm MS}^2}} e^{-\left[\log\psi - \log\psi_{\rm MS}\right]^2/2\sigma_{\rm MS}^2} + \dfrac{\mathcal{N}_{\rm SB}}{\sqrt{2\pi \sigma_{\rm SB}^2}} e^{-\left[\log\psi - \log\psi_{\rm SB}\right]^2/2\sigma_{\rm SB}^2}\,;
    \label{eq:psi_dist}
\end{equation}
where $\mathcal{N}_{\rm MS} = 1 - \mathcal{N}_{\rm SB}$ is the fraction of MS galaxies and $\mathcal{N}_{\rm SB}(\log M_*,z)$ the fraction of SB ones. Following \cite{Chruslinska:2021MS_SB}, we also have
\begin{equation}
    \mathcal{N}_{\rm SB}(\log M_*,z) = 
    \begin{cases}
        0 & \text{ if $\log(M_*)<\log(M_{\rm fix})(z)$}\\
        a_1(z)\log(M_*) + b_1(z) & \text{ if $\log(M_{\rm fix})(z)\leq\log(M_*)\leq\log(M_{\rm 0;SB})(z)$}\\
        a_2(z)\log(M_*) + b_2(z) & \text{ if $\log(M_{\rm 0;SB})(z)<\log(M_*)\leq10.5$}\\
        0.03 & \text{ if $\log(M_*) > 10.5$} \\
    \end{cases}\,,
    \label{eq:N_SB}
\end{equation}
where $(a_1(z), a_2(z), b_1(z), b_2(z), \log(M_{\rm fix})(z), \log(M_{\rm 0;SB})(z))$ is a set of parameters fitted to observational estimates of starburst fractions coming from \cite{Caputi:2017SB_z4} and \cite{Bisigiello:2018SB_z3}, and their values are found in Tab. 2 of \cite{Chruslinska:2021MS_SB}. Eq.~(\ref{eq:N_SB}) holds strictly in the redshift range $0.5<z<4.4$ and for redshifts outside this range the most conservative approximation can be obtained by a linear extrapolation. 

Finally, $\log \psi_{\rm MS}$ in Eq.~(\ref{eq:psi_dist}) is given by the MS fitting from \cite{Popesso:2023MS_SFR}, following a similar approach of Ref. \cite{Speagle:2014logSFR_fit} to fit the MS with time dependent coefficients in the mass range, but with an extra quadratic term in $\log M_*$ motivated by a bending in their most recent MS estimates:
\begin{equation}
    \log \psi_{\rm MS}(\log M_*, z) = (m_0t_{\rm F}(z) + q_0) + (m_1 t_{\rm F}(z) + q_1)\log M_* + q_2\log^2M_*\,;
    \label{eq:galaxy_MS}
\end{equation}
where the parameters $(m_0,m_1, q_0, q_1, q_2)$, listed in the first column of Tab. 2 of \cite{Popesso:2023MS_SFR}, were fitted assuming the fiducial cosmological parameters (F) which also enter into $t_F(z)$ the age of the universe at redshift $z$. We also notice that the MS parametrization in Eq.~(\ref{eq:galaxy_MS}) depends explicitly on $t_{\rm F}(z)$, which in turn is cosmology-dependent. However, Ref. \cite{Speagle:2014logSFR_fit} showed in Sec.~3.2.1 that variations in cosmological parameters induce changes in the inferred MS normalization of less than $\sim 0.05$ dex. Such an effect is well below the intrinsic scatter of the relation and therefore negligible for our purposes. We thus adopt the fiducial cosmology when evaluating Eq.~(\ref{eq:galaxy_MS}). Furthermore, following the results from Fig. 6 in Ref. \cite{Chruslinska:2021MS_SB} we decided to consider $\log \psi_{\rm SB} = \log\psi_{\rm MS} + 0.6$ as a typical location for the SB sequence and the values $\sigma_{\rm MS}\approx0.2$ dex for the MS dispersion and $\sigma_{\rm SB}\approx0.25$ dex for the SB dispersion. 

\begin{figure}[h]
    \centering
    \includegraphics[width=0.7\textwidth]{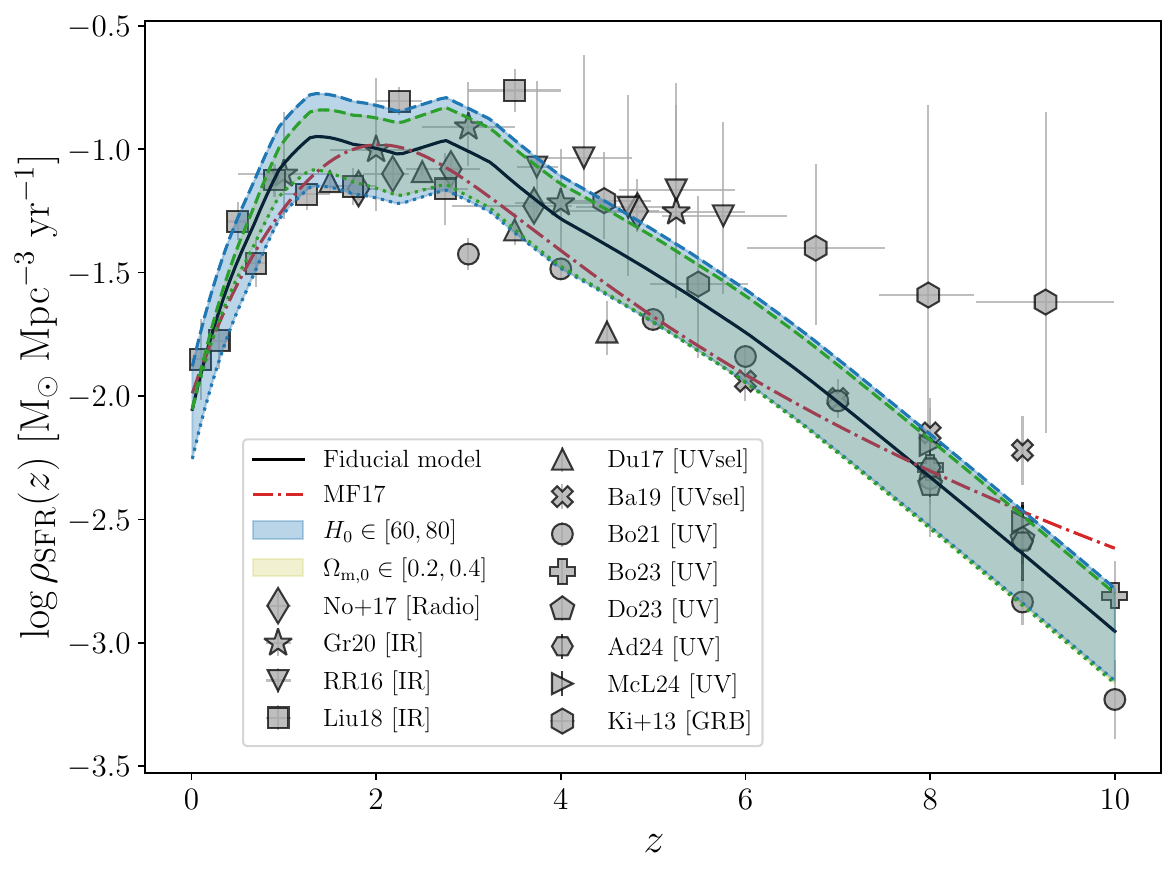}
    \caption{The total cosmic star formation rate density as a function of redshift and different cosmological parameters. Data markers refer to:  (diamonds) \cite{Novak:2017SFR_Radio}, (stars) \cite{Gruppioni:2020SFR_submm}, (inverse triangles) \cite{Rowan-Robinson:2016SFR_IR_highz}, (squares) \cite{Liu:2018SFR_IR}, (triangles) \cite{Dunlop:2017SFR_IR_ALMA}, (crosses) \cite{Bhatawdekar:2019SFR_UV}, (circles) \cite{Oesch:2018SFR_UV, Bouwens:2021SFR_UV}, (plus signs) \cite{Bouwens:2023SFR_UV}, (pentagons) \cite{Donnan:2023SFR_UV}, (hexagons) \cite{Adams:2024SFR_UV}, (rotated triangles) \cite{McLeod:2024SFR_UV}, (rotated hexagons) \cite{Kistler:2013SFR_GRB}. For comparison we also show: the Madau and Fragos SFR density \cite{Madau:2017_SFRD} as the red dot-dashed curve (Eq.~(\ref{eq:MF_SFRD})); and a range of different cosmological parameters, where dotted lines correspond to the lower bound and dashed lines correspond to the upper bound.} 
    \label{fig:SFRD} 
\end{figure}

The resulting SFR density as a function of redshift from Eq.~(\ref{eq:SFRD_tot}) is shown in Fig.~\ref{fig:SFRD} and is also compared with observations from different galaxy surveys from Radio to GRB \cite{Novak:2017SFR_Radio, Gruppioni:2020SFR_submm, Rowan-Robinson:2016SFR_IR_highz, Liu:2018SFR_IR, Dunlop:2017SFR_IR_ALMA, Bhatawdekar:2019SFR_UV, Oesch:2018SFR_UV, Bouwens:2021SFR_UV, Bouwens:2023SFR_UV, Donnan:2023SFR_UV, Adams:2024SFR_UV, McLeod:2024SFR_UV} and with the Madau and Fragos SFR fit (MF17, Eq.~(\ref{eq:MF_SFRD})) \cite{Madau:2017_SFRD}. The SFR density peaks around $z\sim2$ and then declines faster than the MF17 fit, while remaining consistent with high-redshift observations within their uncertainties. The bands show the effect of varying the cosmological parameters.

\begin{figure}[h]
    \centering
    \includegraphics[width=0.6\textwidth]{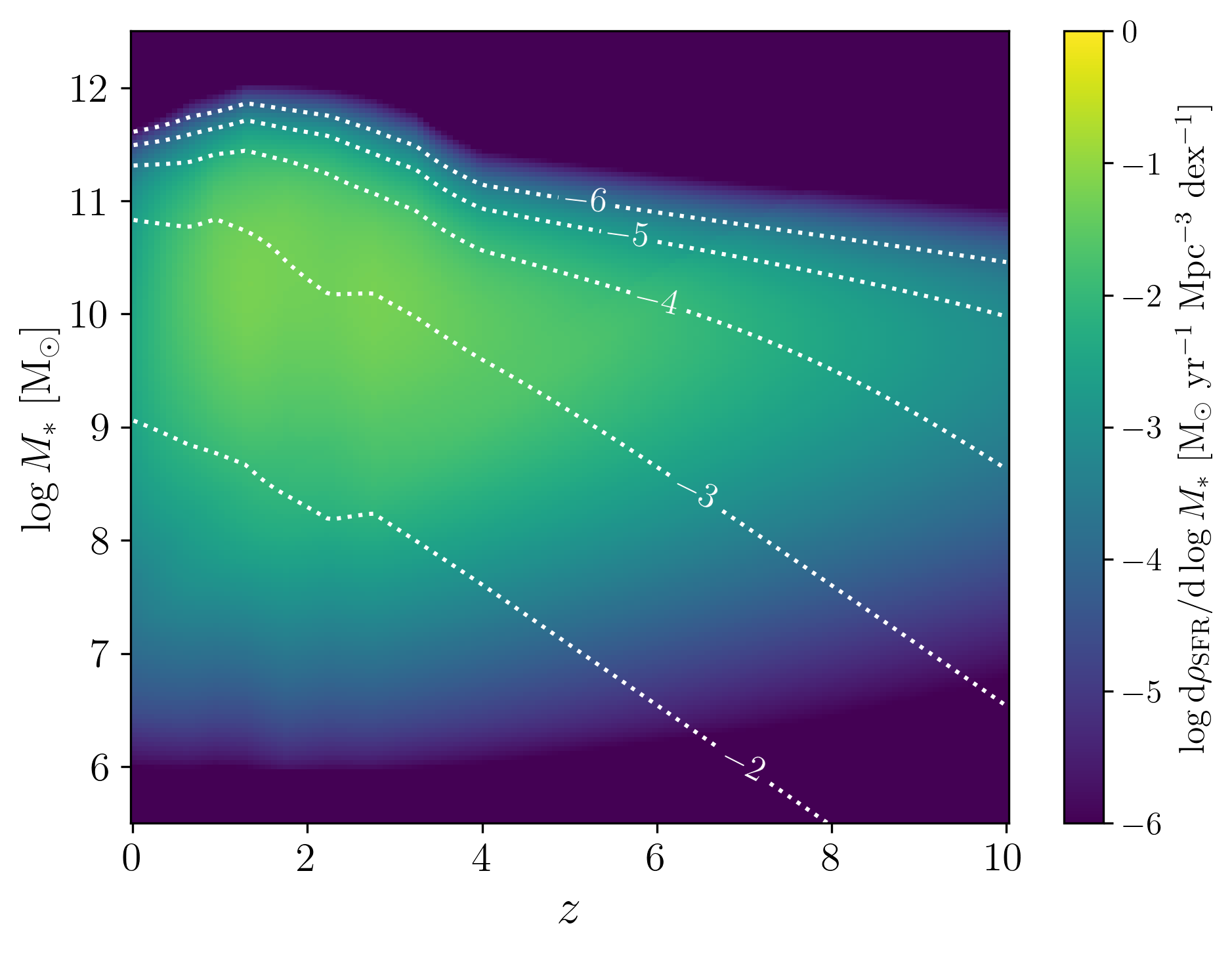}
    \caption{The total cosmic star formation rate density (color-coded) sliced in galaxy stellar mass bins as a function of redshift, i.e. the integrand of Eq.~(\ref{eq:SFRD_tot}) in $\log M_*$, for the fiducial cosmology. The labeled white contours show the log of the stellar mass function for star forming galaxies from Eq.~(\ref{eq:SMF}). We show only the fiducial cosmology, because the effect of varying cosmological parameters is $\lesssim 0.2$ dex (see Sec. \ref{sec:SFRD}) and cannot be appreciated on this color scale.}
    \label{fig:dSFRD_dlogM} 
\end{figure}

The same framework resolves the SFR density in stellar mass, i.e. the integrand of Eq.~(\ref{eq:SFRD_tot}) in $\log M_*$ (Fig.~\ref{fig:dSFRD_dlogM}), which could be used in galaxy-catalog dark-siren analyses \cite{Mastrogiovanni:2023gal_cat}. At $z\sim2$ most stars form in galaxies with $M_*\approx10^{10}$--$10^{11}\,M_\odot$.

The formation efficiency and mass spectrum of compact objects are known to depend sensitively on the metallicity ($Z$) of their progenitor stars (Subsec.~\ref{sec:simulations}). Therefore, to consistently connect star formation to binary black hole merger rates, it is necessary to resolve the SFR density not only in redshift but also in metallicity. We thus compute the SFR density per unit metallicity bin, which will be a central quantity in the merger rate density modeling of Subsec.~\ref{sec:MRD}, linking star formation to binary evolution:
\begin{equation*}
    \dfrac{\mathrm{d}\rho_{\rm SFR}}{\mathrm{d}\log Z}(\log Z, z) = \int \text{d}\log M_* \dfrac{\text{d} N_{\rm SFG}}{\text{d}\log M_*\text{d}V_c}(\log M_*, z)\int \text{d}\log \psi \dfrac{\text{d}p}{\text{d}\log\psi}(\log \psi|\log M_*, z)\times
\end{equation*} 
\begin{equation}
    \times\dfrac{\text{d}p}{\text{d}\log Z}(\log Z|\log M_*, \log \psi, z)\,\psi\,;
    \label{eq:dSFRD_dlogZ}
\end{equation}
where $\text{d}p/\text{d}\log Z$ is the metallicity distribution for fixed mass, SFR and redshift. 

The three observables $(Z, M_*, \psi)$ are known to be tightly correlated in the local Universe through the fundamental metallicity relation (FMR). In this work, we consider two different local calibrations of this relation. The first follows \cite{Curti:2020FMR}: 
\begin{equation}
    12 + \log(O/H) = Z_0 - \dfrac{\gamma}{\beta}\log\left(1+ \left(\dfrac{M_*}{10^{m_0}\psi^{m_1}}\right)^{-\beta}\right)\,,
    \label{eq:FMR_Curti}
\end{equation}
where $Z_0$ represents the asymptotic metallicity at high mass, and $M_0 = 10^{m_0}\psi^{m_1}$ defines a characteristic turnover mass above which the relation saturates.

The second calibration is taken from \cite{Andrews:2013FMR2}:
\begin{equation}
    12 + \log(O/H) = 0.43(\log M_{*} - 0.66 \log\psi) +4.58\,,
    \label{eq:FMR_AM}
\end{equation}
which corresponds to the fit with the lowest intrinsic scatter in Ref. \cite{Andrews:2013FMR2} analysis.  In both cases, we convert oxygen abundances to absolute metallicity using 
\begin{equation}
\log Z = [12+\log(O/H)] - [12+\log(O/H)]_\odot + \log Z_\odot,
\end{equation}
adopting $[12+\log(O/H)]_\odot = 8.76$ \cite{Caffau:2011Z_sun}. 

Recent James Webb Space Telescope's observations \cite{Nakajima:2023FMR_z_10} show that the calibration from \cite{Andrews:2013FMR2} for Eq.~(\ref{eq:FMR_AM}) remains broadly consistent with high-redshift data up to $z\approx 8$, while deviations from the relation of Eq.~(\ref{eq:FMR_Curti}) in \cite{Curti:2020FMR} have been reported for $z \gtrsim 6$. To consistently model the metallicity distribution at high redshift, we incorporate the redshift-dependent corrections inferred in these works. In particular, we apply the evolution measured in \cite{Curti:2023FMR_z} to Eq.~(\ref{eq:FMR_Curti}) (hereafter Cu23) and the high-redshift calibration of \cite{Nakajima:2023FMR_z_10} to Eq.~(\ref{eq:FMR_AM}) (hereafter AM13), thereby constructing explicitly redshift-dependent FMR models.

\begin{figure}[h]
    \centering
    \includegraphics[width=1\textwidth]{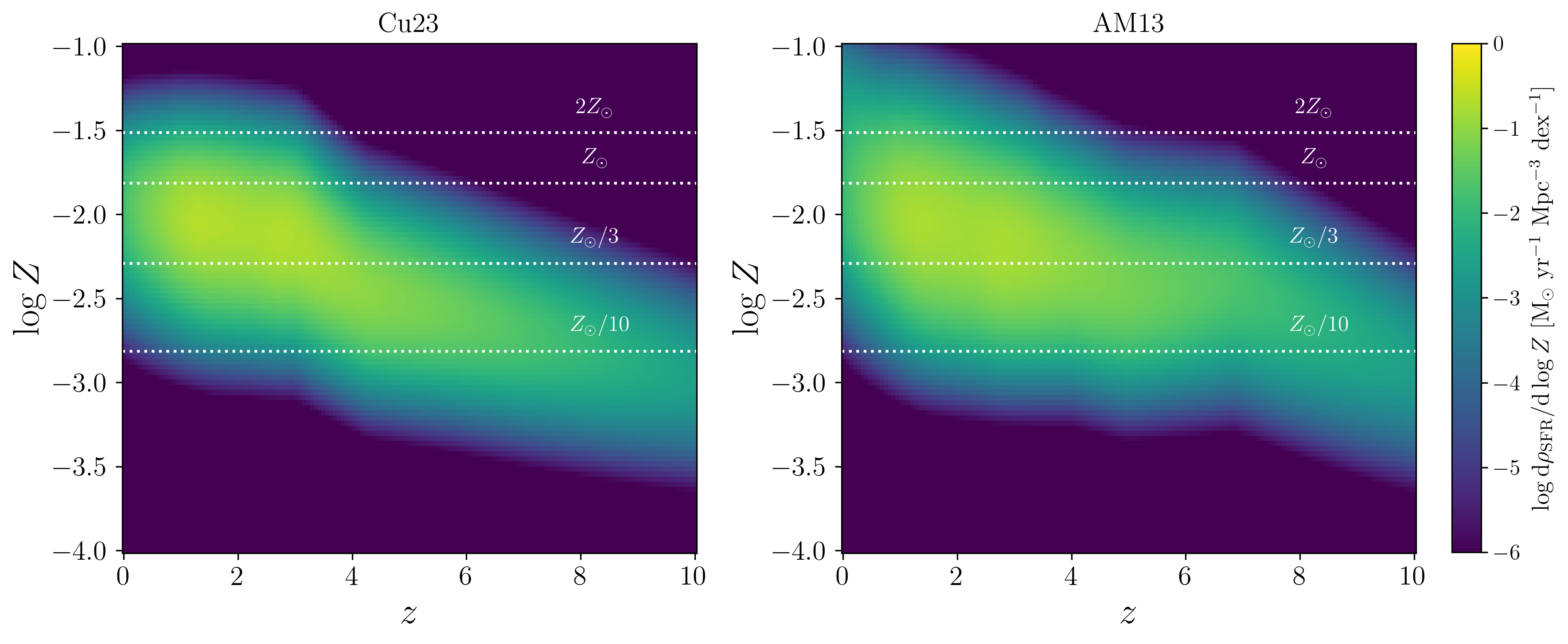}
    \caption{The total cosmic star formation rate density (color-coded) sliced in metallicity bins as a function of redshift, Eq.~(\ref{eq:dSFRD_dlogZ}) for the fiducial cosmology. The two panels refer to the two different FMRs considered, labeled in the text as Cu23 and AM13. In both panels the four horizontal dashed lines highlight the metallicity values of $2Z_{\odot}$, $Z_{\odot}$, $Z_{\odot}/3$, $Z_{\odot}/10$ from top to bottom. Cosmological parameter variations are not presented in these plots being $\lesssim 0.2$ dex (see Sec. \ref{sec:SFRD}) and could not be visible on this color scale.}
    \label{fig:dSFRD_dlogZ} 
\end{figure}

At fixed stellar mass, SFR and redshift, we assume that the metallicity follows a log-normal distribution centered on the value predicted by the FMR, $Z_{\rm FMR}(\log M_*,\log\psi,z)$:
\begin{equation}
    \dfrac{\text{d}p}{\text{d}\log Z}(\log Z|\log M_*,\log \psi,z) = \dfrac{1}{\sqrt{2\pi \sigma_{\rm FMR}^2}}e^{-[\log Z - \log Z_{\rm FMR}]^2/2\sigma^2_{\rm FMR}}\,.
\end{equation}
We adopt a dispersion $\sigma_{\rm FMR} \simeq 0.15$ dex, representative of the intrinsic scatter observed in both local and high-redshift galaxy samples \cite{Curti:2020FMR, Curti:2023FMR_z, Andrews:2013FMR2, Nakajima:2023FMR_z_10}. Fig.~\ref{fig:dSFRD_dlogZ} compares the two FMRs. In both cases star-forming gas was more metal-poor in the past. At low redshift AM13 gives a broader metallicity distribution than Cu23, so a wider range of metallicities contributes to the merger rate.

\subsection{Mapping stellar birth to binary mergers}\label{sec:MRD}
In this section, we construct the binary black hole (BBH) merger rate density by combining the total SFR density derived in Sec.~\ref{sec:SFRD} with the outputs of isolated binary population synthesis simulations. Our goal is a coherent merger-rate framework for the number-count analysis, not a comparison between binary-evolution scenarios.

\subsubsection{Population synthesis simulations}\label{sec:simulations}
Several formation channels likely contribute to the observed BBH population \cite{Zevin:2021pop_synth_4, Cheng:2023GW_and_channels, Spera:2022GW_binary_review}. Here we consider only isolated binaries, with a representative set of prescriptions. This choice does not affect the internal consistency of the forecast, because the same model generates and analyzes the mock data, but we will show that it matters when we will compare our results with GWTC-5.0 in Sec.~\ref{sec:discussion}.

We run the simulations with the population-synthesis code {\footnotesize\textup{\textrm{SEVN}}} (Stellar EVolution for N-body) \cite{Spera:2017sevn_1, Spera:2019sevn_2, Iorio:2023sevn_4}, version 2.10.4\footnote{{\scriptsize\textup{\textrm{SEVN}}} is publicly available at \href{https://gitlab.com/sevncodes/sevn}{https://gitlab.com/sevncodes/sevn}.}. {\footnotesize\textup{\textrm{SEVN}}} evolves single stars by interpolating pre-computed stellar tracks and treats binary interactions with analytic prescriptions. We use {\footnotesize\textup{\textrm{PARSEC}}} tracks \cite{Bressan:2012parsec_1, Costa:2019parsec_2, Costa:2019parsec_3, Nguyen:2022parsec_4} with overshooting parameter $\lambda_{\rm ov}=0.5$ for H-rich stars, and pure-He tracks for stripped stars. For compact-object formation we adopt the \emph{delayed} core-collapse model \cite{Fryer:2012delayed}, the \emph{mapelli20} prescription for pair instability \cite{Spera:2017sevn_1, Mapelli:2020sevn_3}, the \emph{unified} natal kicks \cite{Giacobbo:2020natal_kick} and a maximum neutron-star mass of $3\,M_\odot$, which is also the minimum BH mass in our catalogs. Mass-transfer stability follows the default critical mass ratios of {\footnotesize\textup{\textrm{SEVN}}}, based on \cite{Hurley:2002q_crit}.

Unstable mass transfer can lead to a common-envelope (CE) phase \cite{Ivanova:2013CE}. {\footnotesize\textup{\textrm{SEVN}}} treats it with the standard $\alpha\lambda$ formalism \cite{Webbink:1984alpha_lambda, Hurley:2002q_crit}: a fraction $\alpha_{\rm CE}$ of the orbital energy released during the inspiral is used to unbind the envelope,

\begin{equation}
    E_{\rm bind} = \alpha_{\rm CE} \Delta E_{\rm orb}\,,
\end{equation}
where the binding energy of the envelope depends on the stellar structure through the parameter $\lambda$ \cite{Claeys:2014lambdas}. Larger values of $\alpha_{\rm CE}$ make envelope ejection easier and leave wider post-CE binaries. Given the large uncertainties on CE evolution, we run six sets of simulations with different $\alpha_{\rm CE}$ (Tab.~\ref{tab:sim_pars}).

For each $\alpha_{\rm CE}$ and metallicity we evolve $10^7$ binaries, with the initial distributions listed in Tab.~\ref{tab:sim_pars} and no initial rotation. We sample primaries above $8\,M_\odot$ and secondaries above $5\,M_\odot$ to improve the sampling of BBH progenitors, and correct for these cuts with $f_{\rm IMF}=0.185$, the fraction of the total mass of a full Kroupa population that falls in the simulated range. Each set corresponds to $M_{\rm tot,\star\star}=3.9\times10^8\,M_\odot$, and we simulate 27 metallicities uniformly spaced in $\log Z\in[-4,-1.4]$.

\begin{table}[htbp]
\centering
\begin{tabular}{|c|c|}
\hline
\textbf{Parameters and prescriptions}& \textbf{Values}\\
\hline
$\alpha_{\rm CE}$ & $0.5, 1, 2, 3, 4, 5$\\
$\log Z$ & $27$ values linearly spaced in $[-4, -1.4]$ \\
simulated binaries $\forall\alpha_{\rm CE}$& $27\times 10^{7}$\\
SN mechanism & \emph{delayed}\\
PISN & \emph{mapelli20}\\
natal kick & \emph{unified}\\
primary mass $m_{\rm ZAMS,1}$ & $p\propto m_{\rm ZAMS,1}^{-2.3}$, $[8,150]\,M_\odot$\\
mass ratio $q$ & $p\propto q^{-0.1}$, $q\geq\max(5\,M_\odot/m_{\rm ZAMS,1},0.1)$\\
orbital period & $p\propto\mathcal{P}^{-0.55}$, $\mathcal{P}=\log(P/{\rm day})\in[0.15,5.5]$\\
eccentricity & $p\propto e^{-0.42}$, $e\in[0,0.9]$\\
\hline
\end{tabular}
\caption{Initial conditions, prescriptions and parameters of our simulations. Initial mass function from \cite{Kroupa:2001IMF, Weidner:2004IMF_lim1, Shenar:2023IMF_lim2}, mass ratio, period and eccentricity distributions from \cite{Sana:2012initcond}.}
\label{tab:sim_pars}
\end{table}

\subsubsection{BBH merger rate density model}\label{sec:MRD_detail}

We model the BBH merger rate density by convolving the cosmic star formation rate density (resolved in metallicity) with the outcomes of our population-synthesis simulations. The simulations provide, at each progenitor metallicity $Z$ and common-envelope efficiency $\alpha_{\rm CE}$, both i) the overall number of merging BBHs produced per unit stellar mass formed and ii) the distribution of BBH intrinsic parameters and delay times (time between star binary formation and GW merger).

We define the merger distribution function as the number of BBH mergers produced per unit stellar mass formed, per unit intrinsic parameter space,

\begin{equation}
\frac{\mathrm{d}N_{\star\star\rightarrow\bullet\bullet\rightarrow\bullet}}{\mathrm{d}M_{\star\star}\,\mathrm{d}q\,\mathrm{d}\mathcal{M}_{\rm s}\,\mathrm{d}\log\tau}=\eta(\log Z,\alpha_{\rm CE})\frac{\mathrm{d}p}{\mathrm{d}q\,\mathrm{d}\mathcal{M}_{\rm s}\,\mathrm{d}\log\tau}(q,\mathcal{M}_{\rm s}, \log\tau| \log Z,\alpha_{\rm CE}),
\label{eq:merg_dist}
\end{equation}
where the conditional probability density encodes the shape of the distribution in mass ratio $q$, source-frame chirp mass $\mathcal{M}_{\rm s}$, and delay time $\tau$. The overall normalization is set by the merger efficiency $\eta(\log Z,\alpha_{\rm CE})$, which is defined as the number of BBH mergers produced per unit total stellar mass formed at log metallicity $\log Z$ \cite{Mapelli:2017merg_eff_1, Giacobbo:2018merg_eff_2, Giacobbo:2018merg_eff_3}, i.e.

\begin{equation}
\eta(\log Z,\alpha_{\rm CE}) \equiv f_{\rm IMF}\, f_{\rm bin}\,\frac{N_{\tau<t_{\rm age}}(\log Z,\alpha_{\rm CE})}{M_{\rm tot,\star\star}},
\label{eq:merg_eff}
\end{equation}
where $N_{\tau<t_{\rm age}}$ is the number of simulated systems that form a BBH and merge within the age of the Universe $t_{\rm age}$, $M_{\rm tot,\star\star}$ is the total simulated stellar mass at that metallicity, $f_{\rm bin}=0.5$ accounts for the binary fraction in isolated systems, and $f_{\rm IMF}$ corrects for the restricted IMF sampling adopted in the simulations. Furthermore, by definition, the merger efficiency counts systems that merge within the age of the Universe. Since the latter depends on the assumed cosmological parameters, $\eta(\log Z,\alpha_{\rm CE})$ inherits a corresponding cosmological dependence.

\begin{figure}[h]
     \centering
     \includegraphics[width=0.7\textwidth]{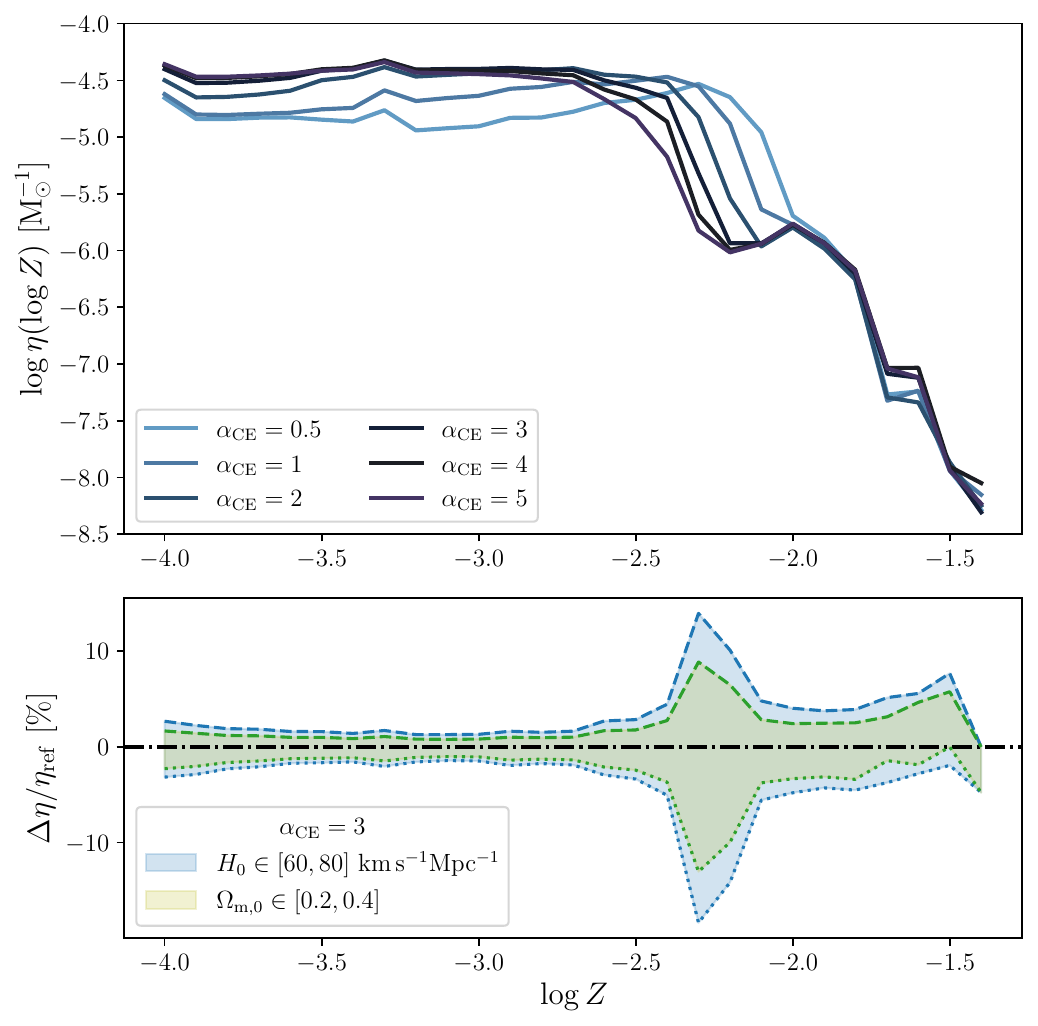}
     \caption{Merger efficiency $\eta(\log Z,\alpha_{\rm CE})$ (Eq.~(\ref{eq:merg_eff})) calculated from the {\footnotesize\textup{\textrm{SEVN}}} simulations (Subsec. \ref{sec:simulations}) and adopting the age of the Universe corresponding to the fiducial cosmological parameters reported in Sec.~\ref{sec:SFRD}, $t_{\rm age}^{F} = 13.5$ Gyr. \emph{Top panel:} $\log\eta(\log Z,\alpha_{\rm CE})$ as a function of $\log Z$ for different values of the common-envelope efficiency $\alpha_{\rm CE}$, as indicated in the legend. \emph{Bottom panel:} relative variation of the merger efficiency, $\Delta \eta/\eta_{\rm ref}$, with respect to the reference value $\eta_{\rm ref}(\log Z)$ for $\alpha_{\rm CE}=3$. The blue shaded region shows the variation of $\eta(Z,\alpha_{\rm CE}=3)$ obtained by varying $H_0$ in the interval $[60,80]\,{\rm km\,s^{-1}\,Mpc^{-1}}$ at fixed fiducial $\Omega_{\rm m,0}$; while the green shaded region shows the range obtained by varying $\Omega_{\rm m,0}$ in $[0.2,0.4]$ at fixed fiducial $H_0$. For each band, dashed and dotted curves mark the lower and upper cosmological parameter bounds, respectively.}
     \label{fig:merg_eff} 
 \end{figure}

Figure~\ref{fig:merg_eff} shows the behavior of the merger efficiency for our set of simulations (Subsec.~\ref{sec:simulations}). The top panel shows $\eta(\log Z,\alpha_{\rm CE})$ as a function of metallicity for different values of $\alpha_{\rm CE}$, assuming the fiducial age of the Universe, $t^{\rm F}_{\rm age} = 13.5$ Gyr. The merger efficiency decreases steeply with increasing metallicity, spanning about four orders of magnitude, while $\alpha_{\rm CE}$ mainly shifts the metallicity at which this drop occurs and changes the low-metallicity plateau by $\lesssim 0.5$ dex.

The bottom panel quantifies the cosmological dependence of $\eta$ at fixed $\alpha_{\rm CE}=3$. The variation is within a few percent at $\log Z \lesssim -2.6$ and reaches $\approx +15$/$-20\%$ around $\log Z\approx-2.3$, where $\eta$ drops steeply and a larger fraction of merging systems has delay times comparable to $t_{\rm age}$. This dependence cancels in the merger rate, because the conditional distribution in Eq.~(\ref{eq:merg_dist}) is normalized over the same systems with $\tau < t_{\rm age}$, i.e. 
\begin{equation}
    \frac{\mathrm{d}p}{\mathrm{d}q\,\mathrm{d}\mathcal{M}_{\rm s}\,\mathrm{d}\log\tau}\approx \dfrac{N_j}{N_{\tau<t_{\rm age}}\Delta q\Delta\mathcal{M}_{\rm s}\Delta \log\tau}\,,
\end{equation}
with $N_{j}$ the systems in the j-th $\Delta q \Delta\mathcal{M}_{\rm s}\Delta \log\tau$ bin, and thus the product with $\eta$ cancels the cosmological dependence.

In the full simulation output, $(q,\mathcal{M}_{\rm s},\tau)$ are correlated (e.g., through binary interactions that affect both the final masses and the orbital separation, and hence the inspiral time). Since these correlations can impact both the shape and the low-redshift normalization of the predicted merger rate density, we explicitly test how sensitive our results are to some simplifying assumptions. We consider three alternative \emph{factorizations}:

\begin{equation}
    \dfrac{\text{d}p}{\text{d}q\,\text{d}\mathcal{M}_{\rm s}\text{d}\log \tau}= 
    \begin{cases}
    \text{No factorization}&\text{F0}\\
    \dfrac{\text{d}p}{\text{d}q\,\text{d}\mathcal{M}_{\rm s}}(q,\mathcal{M}_{\rm s}|\log Z,\alpha_{\rm CE}) \dfrac{\text{d}p}{\text{d}\log\tau}(\log \tau|\log Z,\alpha_{\rm CE}) & \text{F1} \\
    C(\alpha_{\rm CE})\dfrac{\text{d}p}{\text{d}q\,\text{d}\mathcal{M}_{\rm s}}(q,\mathcal{M}_{\rm s}|\log Z,\alpha_{\rm CE}) &\text{F2}
\end{cases}\,\,\,\,\,,
\label{eq:factorization}
\end{equation}
with the distributions computed on a fixed grid $(q, \mathcal{M}_{\rm s}, \log\tau)$ for each $\alpha_{\rm CE}$, common to all metallicities, uniformly spaced in $q \in [q_{\rm min}(\alpha_{\rm CE}), 1]$, in $\mathcal{M}_{\rm s} \in [\mathcal{M}_{\rm s,min}(\alpha_{\rm CE}), \mathcal{M}_{\rm s,max}(\alpha_{\rm CE})]$ and in $\log\tau \in [\log\tau_{\rm min}(\alpha_{\rm CE}), \log t_{\rm age}]$, where the lower and upper bounds of $q$ and $\mathcal{M}_{\rm s}$, and the lower bound of $\log\tau$, are the extreme values found across all metallicities of the given $\alpha_{\rm CE}$ simulation. The factorizations are divided into three cases: i) F0, retaining all correlations coming from the simulations; ii) F1, assuming that the delay time is statistically independent of $(q,\mathcal{M}_{\rm s})$ for all $(\log Z,\alpha_{\rm CE})$; and iii) F2, which further assumes a log-flat delay-time distribution, so that $\mathrm{d}p/\mathrm{d}\log\tau$ reduces to the normalization constant $C(\alpha_{\rm CE}) = [\log t_{\rm age} - \log\tau_{\rm min}(\alpha_{\rm CE})]^{-1}$. 

We then convolve the merger distribution function (Eq.~\ref{eq:merg_dist}) with the star formation rate density per unit metallicity, $\mathrm{d}\rho_{\rm SFR}/\mathrm{d}\log Z$ (Eq.~\ref{eq:dSFRD_dlogZ}). A system merging at redshift $z$ with delay time $\tau$ must have formed at cosmic time $t(z)-\tau$, corresponding to a formation redshift $\tilde{z}$ defined implicitly by $t(\tilde{z}) = t(z)-\tau$; delay times with $\tau > t(z)$, which would imply formation before the Big Bang, do not contribute and are removed from the computation. The differential merger rate density in the source frame $(t_{\rm s})$ is therefore

\begin{equation*}
    \dfrac{\text{d}N}{\text{d}V_c\text{d}t_{\rm s}\text{d}q\,\text{d}\mathcal{M}_{\rm s}}(z, q, \mathcal{M}_{\rm s}) = \int \text{d}\log Z \int\text{d}\log \tau\,\dfrac{\text{d}\rho_{\rm SFR}}{\text{d}\log Z}(\log Z,\tilde{z}(t(z)-\tau))\times
\end{equation*}
\begin{equation}
    \times\eta(\log Z, \alpha_{\rm CE}) \dfrac{\text{d}p}{\text{d}q\,\text{d}\mathcal{M}_{\rm s}\text{d}\log \tau}(q, \mathcal{M}_{\rm s},\log \tau|\log Z, \alpha_{\rm CE})\,;
    \label{eq:diff_MRD}
\end{equation}
and the merger rate density redshift evolution can be obtained by integrating over $q$ and $\mathcal{M}_{\rm s}$:

\begin{equation}
    \mathcal{R}(z) = \dfrac{\text{d}N}{\text{d}V_c\text{d}t_{\rm s}}(z) = \int\text{d}q\,\text{d}\mathcal{M}_{\rm s}\dfrac{\text{d}N}{\text{d}V_c\text{d}t_{\rm s}\text{d}q\,\text{d}\mathcal{M}_{\rm s}}\,.
    \label{eq:MRD}
\end{equation}

Since in F1 the delay-time distribution is the marginal of the F0 one, F0 and F1 give exactly the same mass-integrated rate, $\mathcal{R}_{\rm F0}(z)=\mathcal{R}_{\rm F1}(z)$. We keep F1 in the analysis because the two differ in the joint distribution of masses and redshift, which affects the detectable fraction.

\begin{figure}[h]
    \centering
    \includegraphics[width=1\textwidth]{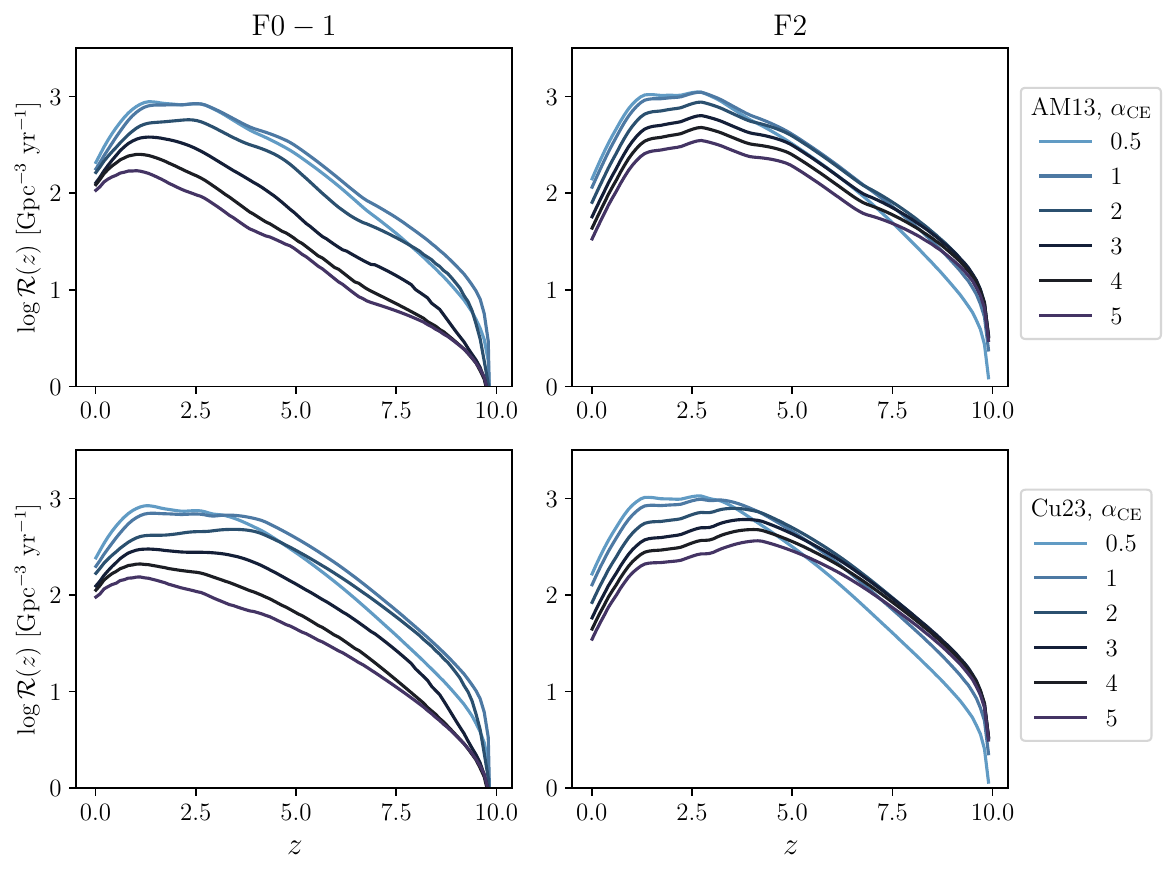}
    \caption{Logarithm of the merger rate density $\log \mathcal{R}(z)$, as a function of redshift $z$ for fiducial cosmological parameters. The left column shows F0 and F1, which give the same mass-integrated rate; the right column shows F2 (Eq.~(\ref{eq:factorization})). Rows correspond to the two FMR prescriptions (AM13, top; Cu23, bottom) and colors to different values of $\alpha_{\rm CE}$, as indicated in the legend. }
    \label{fig:MRD_comp} 
\end{figure}

In Fig. \ref{fig:MRD_comp} we compare the merger rate densities obtained for $\alpha_{\rm CE}\in[0.5,5]$, for the two FMR prescriptions introduced in Subsec.~\ref{subsec:SFRD_model}, and for the three merger distribution function factorizations defined in Eq.~(\ref{eq:factorization}). We find that choosing the F(0,1) or F2 factorization significantly impacts both the local rate $\mathcal{R}(z=0.2)=\mathcal{R}_{0.2}$ and the redshift evolution of the merger rate density, apart from the common drop-off around $z=10$ as a direct consequence of low star formation from the cosmic SFR. The F2 model, based on a log-flat delay-time distribution, produces a different behavior with respect to F(0,1): it predicts a downward shift in the local rate, by 26.5--61.1\% depending on $\alpha_{\rm CE}$ and the FMR, while simultaneously enhancing the high-redshift tail, resulting in a flatter redshift evolution. In Tab. \ref{tab:local_rates} we present the comparison for the predictions on the local merger rates with all the models at different $\alpha_{\rm CE}$.

\begin{table}[h]
\centering
\begin{tabular}{|llcccccc|}
\hline
FMR & Factorization & $\alpha_{\rm CE}=0.5$ & $\alpha_{\rm CE}=1$ & $\alpha_{\rm CE}=2$ & $\alpha_{\rm CE}=3$  & $\alpha_{\rm CE}=4$  & $\alpha_{\rm CE}=5$  \\
\hline
AM13 & F0--F1 & 301 & 244 & 215 & 170 & 159 & 131 \\
AM13 & F2     & 218 & 177 & 122 &  86 &  65 &  51 \\
Cu23 & F0--F1 & 343 & 264 & 214 & 160 & 143 & 116 \\
Cu23 & F2     & 252 & 194 & 125 &  86 &  66 &  52 \\
\hline
\end{tabular}
\caption{Local merger rate density $\mathcal{R}_{0.2}=\mathcal{R}(z=0.2)$ $[\text{Gpc}^{-3}\text{yr}^{-1}]$ for the two FMR prescriptions (AM13, Cu23), the factorizations of Eq.~(\ref{eq:factorization}) (F0 and F1 coincide for the mass-integrated rate, see Eq.~\ref{eq:MRD}) and the six simulated values of $\alpha_{\rm CE}$, at the fiducial cosmology.}
\label{tab:local_rates}
\end{table}

The choice of FMR matters less. For F0 and F1, Cu23 gives higher local rates than AM13 for $\alpha_{\rm CE}\leq1$ and lower ones for larger $\alpha_{\rm CE}$, with differences of at most 15\% (Tab.~\ref{tab:local_rates}). Cu23 also gives a broader plateau around the peak of the merger rate.

In our work, we adopt the AM13 F0 model as our fiducial scenario. This choice combines the FMR most consistent with current data and the full merger distribution function without simplifying factorizations, thereby preserving the correlations among intrinsic parameters and delay time predicted by the {\footnotesize\textup{\textrm{SEVN}}} simulations. In Section \ref{sec:pop_synth_uncertain} we further investigate the fiducial model and compare the predicted and observed local merger rates.

\subsection{Cosmological and astrophysical inference from BBH number counts}\label{sec:analysis}
In this section, we connect our astrophysical modeling to GW population analysis, focusing only on the contribution from number counts of BBH merger events.
\subsubsection{Hierarchical inference}

The statistical properties of the BBH population are described by the differential merger rate function, defined as the number of merger events per unit detector-frame time, per unit redshift, and per unit intrinsic parameter space. In our framework, the merger rate depends on a set of hyper-parameters $\Lambda=(H_0, \alpha_{\rm CE})$, which encode, respectively, the cosmological expansion rate and the key astrophysical parameter controlling common-envelope evolution. Moreover, we fix $\Omega_{\rm m,0}$ to its fiducial value in the analysis since independent probes are expected to constrain it at the percent level by the ET era, and number counts alone are known to exhibit strong degeneracies between $H_0$ and $\Omega_{\rm m,0}$ \cite{Antinozzi:2024_number_paper}, which would obscure the cosmo-astrophysical degeneracy between $H_0$ and $\alpha_{\rm CE}$ that we aim to characterize.

At the event level, we use  $\theta=(z,q,\mathcal{M}_{\rm s})$ to denote the true source parameters: redshift, mass ratio, and source-frame chirp mass. However, these are not directly observed. Instead, gravitational-wave detectors measure 
a combination of the source-frame parameters, namely the detector-frame chirp mass $\mathcal{M}_{\rm d}=(1+z)\mathcal{M}_{\rm s}$ and luminosity distance $d_L(z)$, in addition to other binary parameters that are inferred from noisy data $x$ in the form of posterior distributions.

The goal of hierarchical Bayesian inference is to constrain the population hyper-parameters $\Lambda$ using a catalog of observed events $\left\{ x_i\right\}^{N_{\rm obs}}_{i=1}$. This requires linking three ingredients: (i) the population model describing the distribution of source parameters under $\Lambda$, (ii) the single-event likelihood $\mathcal{L}_{\rm GW}(x_i|\theta,\Lambda)$, and (iii) the selection function accounting for detectability.

To obtain the differential merger rate from the differential merger rate density of Eq.~(\ref{eq:diff_MRD}) we transform from comoving volume to redshift using 
\begin{equation}
    \dfrac{\text{d}V_c}{\text{d}z}(z) = \dfrac{4\pi c}{H(z)} \dfrac{d_L^2(z)}{(1+z)^2}
\end{equation}
and from source-frame time to detector frame with a factor $(1+z)^{-1}$ due to the time dilation
\begin{equation}
    \dfrac{\text{d}N}{\text{d}t_{\rm d}\,\text{d} z\,\text{d}q\,\text{d}\mathcal{M}_{\rm s}\,} = \dfrac{\text{d}N}{\text{d}V_c\text{d}t_{\rm s}\text{d}q\,\text{d}\mathcal{M}_{\rm s}}\dfrac{\text{d}V_c}{\text{d}z} \dfrac{1}{1+z}\,,
    \label{eq:diff_MR}
\end{equation}
this merger rate fully specifies the population model. 

For convenience, we define the normalized population distribution as
\begin{equation}
\pi_{\rm pop}(\theta|\Lambda)=
\frac{T_{\rm obs}}{N_{\rm id}(\Lambda)}
\frac{\mathrm{d}N}{\mathrm{d}t_{\rm d}\,\mathrm{d}\theta}(\theta,\Lambda),
\label{eq:pop_model}
\end{equation}
where

\begin{equation}
N_{\rm id}(\Lambda)=T_{\rm obs}\int \mathrm{d}\theta\,
\frac{\mathrm{d}N}{\mathrm{d}t_{\rm d}\,\mathrm{d}\theta}(\theta,\Lambda)
\end{equation}

is the total number of BBH merger events observed by an \emph{ideal} detector in the time $T_{\rm obs}$ fixed to $1$ yr in the analysis (assuming the merger rate roughly constant over detector time) and this quantity also acts as a normalizing constant for Eq.~(\ref{eq:pop_model}). Thus, $\pi_{\rm pop}(\theta|\Lambda)$ represents the normalized distribution of source parameters predicted by the merger rate model.

The expected number of \emph{detectable} events during an observing time $T_{\rm obs}$ is then

\begin{equation}
N_{\rm det}(\Lambda)= T_{\rm obs}
\int
\mathrm{d}\theta\,
p_{\rm det}(\theta,\Lambda)
\frac{\mathrm{d}N}{\mathrm{d}t_{\rm d}\,\mathrm{d}\theta}(\theta,\Lambda),
\label{eq:N_det}
\end{equation}

where 
\begin{equation}
    p_{\rm det}(\theta,\Lambda) = p(\text{detectable}|\theta,\Lambda) \equiv \int_{x\in\text{detectable}} \mathcal{L}_{\rm GW}(x|\theta, \Lambda) \text{d} x\,,
    \label{eq:pdet}
\end{equation}
is the detection probability \cite{Vitale:2022hier_like1, Mandel:2019hier_like2}, i.e. the probability that a source with true parameters $\theta$ produces data above the detection threshold. It encodes the selection effects.

Given a catalog of $N_{\rm obs}$ observed events ${x_i}$, the full hierarchical likelihood for the hyper-parameters $\Lambda$ is given by \cite{Vitale:2022hier_like1, Mandel:2019hier_like2}:
\begin{equation}
\mathcal{L}(x_1,...,x_{N_{\rm obs}}, N_{\rm obs}|\Lambda)
\propto
e^{-N_{\rm det}(\Lambda)}
N_{\rm det}(\Lambda)^{N_{\rm obs}}
\prod_{i=1}^{N_{\rm obs}}
\frac{
\int \mathrm{d}\theta\,
\mathcal{L}_{\rm GW}(x_i|\theta,\Lambda)
\pi_{\rm pop}(\theta|\Lambda)
}{
\int \mathrm{d}\theta\,
p_{\rm det}(\theta,\Lambda)
\pi_{\rm pop}(\theta|\Lambda)
}.
\label{eq:hier_like}
\end{equation}

The prefactor accounts for the Poisson probability of observing $N_{\rm obs}$ events when $N_{\rm det}(\Lambda)$ are expected, while the product accounts for the distribution of their intrinsic parameters under the population model and selection effects.

Integrating Eq.~(\ref{eq:hier_like}) over the data $x_i$ of each detectable event, each factor of the product integrates to unity because of Eq.~(\ref{eq:pdet}), and only the Poisson term survives:

\begin{equation}
\mathcal{L}(N_{\rm obs}|\Lambda) \propto e^{-N_{\rm det}(\Lambda)} N_{\rm det}(\Lambda)^{N_{\rm obs}}\,.
    \label{eq:number_like}
\end{equation}

The marginalized likelihood of Eq.~(\ref{eq:number_like}), from now on called \emph{number likelihood}, is a well-defined likelihood for $\Lambda$, although it retains only the information carried by the total number of detections and discards that carried by the distribution of individual event parameters. More precisely, this approach marginalizes over GW data but it still conditions on the fact that we have detected the GW events; a derivation is given in App. \ref{app:marg_like} (see also App. I of \cite{Colombo:2026xkj}). We therefore refer to this as an information-reduced hierarchical analysis, as it yields a well-defined posterior, but one that cannot break degeneracies which the full likelihood of Eq.~(\ref{eq:hier_like}) would resolve. 

Moreover, we argue that although the number likelihood has reduced information if compared to the full hierarchical likelihood, it still has some advantages. First, it is computationally more efficient to calculate and therefore it can be used to analyze a large number of GW events (e.g. our fiducial number of detections $\sim 10^5$ for a single year with ET, see Sec. \ref{sec:results}) that could not be analyzed with the full hierarchical likelihood \cite{Tagliazucchi:2026dpr}. 
Second, it does not suffer from the numerical instabilities that the full hierarchical likelihood has, which are related to the finite nature of the GW posterior samples. For completeness, we will defer the full analysis to a follow-up paper in preparation.

\subsubsection{BBH number counts}\label{sec:BBH_NC}

The detection probability $p_{\rm det}$ is not available in analytical form, and we therefore estimate the integral in Eq.~(\ref{eq:N_det}) through an approximation using a Monte Carlo procedure based on GW signal injections \cite{Abbott:2021GW_pop}. In our analysis, the intrinsic parameters entering the merger rate model are $\theta=(z,q,\mathcal{M}_{\rm s})$ while the remaining quantities needed to generate a GW signal are treated as nuisance parameters. For non-spinning BBHs, these are the inclination angle $\theta_{\rm JN}$, right ascension R.A., declination $\delta$, polarization angle $\psi$, phase at coalescence $\phi_c$ and coalescence time $t_c$. The luminosity distance $d_L$ and detector-frame chirp mass are obtained from the sampled intrinsic parameters through $d_L=d_L(z,H_0)$ and $\mathcal{M}_{\rm d} = \mathcal{M}_{\rm s}(1+z)$. The priors adopted for the injections are summarized  in Tab.~\ref{tab:priors}. In particular, the intrinsic parameters $(z,q,\mathcal{M}_{\rm s})$ are drawn from a common injection prior $\pi_{\rm inj}(\theta|\Lambda)$, while the angular variables are sampled independently and from the distributions listed in the table.

\begin{table}[htbp]
\centering
\begin{tabular}{|c|c|}
\hline
\textbf{BBH parameters}& \textbf{Priors}\\
\hline
$(z, q, \mathcal{M}_{\rm s})$ & $\pi_{\rm inj}(\theta|\Lambda)$\\
$\theta_{\rm JN}$ & $\mathcal{U}_{\cos\theta_{\rm JN}}(-1, 1)$\\
R.A. & $\mathcal{U}_{\text{R.A.}}(0, 2\pi)$\\
$\delta$ & $\mathcal{U}_{\,\sin\delta}(-1, 1)$\\
$\psi$ & $\mathcal{U}_{\,\psi}(0, \pi)$\\
$\phi_c$ & $\mathcal{U}_{\,\phi_c}(0, 2\pi)$\\
$t_c$ &  $\mathcal{U}_{\,t_c}(01/01/2030, 12/31/2030)$\\
\hline
\end{tabular}
\caption{Summary of the priors for GW injections}
\label{tab:priors}
\end{table}

To generate the injections we use {\footnotesize\textup{\textrm{GWFISH}}}, a software package for compact-binary GW simulations based on the Fisher approximation \cite{Dupletsa:2023GWFish}. In the present context, however, we only use the part of the code that computes the noise-free network signal-to-noise ratio (SNR)
\begin{equation}
    \rho_{\omega}^2 = 4\text{Re}\left[\int_0^{\infty}\text{d}f\,\,\dfrac{|h_{\omega}(f)|^2}{S_n(f)}\right]\,,
\end{equation}
where $h_{\omega}(f)$ is the frequency-domain waveform evaluated at the binary parameters $\omega$, and $S_n(f)$ is the detector noise power spectral density. We adopt the \texttt{IMRPhenomD} waveform model \cite{Husa:2016IMRPhenomD1, Khan:2016IMRPhenomD2} and the ET-D sensitivity curve of the Einstein Telescope triangular configuration, i.e. a network of three nested interferometers with 10 km arm length \cite{Maggiore:2020ETD, Abac:2025BlueBook}. We impose a network detection threshold $\rho_{\omega}>\rho_{\rm thr}=12$\footnote{Since we consider the triangular ET configuration, the network SNR is computed as $\rho_{\omega} = \sqrt{\rho_{\rm ET1}^2 + \rho_{\rm ET2}^2 + \rho_{\rm ET3}^2}$. In a simplified case of equal SNR across the three arms, then the corresponding threshold for each individual interferometer is therefore $\rho_{\rm thr, ETi}\approx7$.}.  

Operationally, the detection probability entering Eq.~(\ref{eq:N_det}) is estimated by sampling the binary parameters through the injection campaign. For each simulated source, we evaluate whether the injected signal satisfies the detection threshold, and we retain only the injections that do. In this way, the integral for $N_{\rm det}$ can be approximated by importance sampling over the injection prior:
\begin{equation}
    N_{\rm det}(\Lambda)\approx \dfrac{T_{\rm obs}}{N_{\rm gen}} \sum_{j=1}^{N_{\rm thr}} \dfrac{1}{\pi_{\rm inj}(\theta_j|\Lambda)}\dfrac{\text{d}N}{\text{d}t_{\rm d}\text{d}\theta}(\theta_j|\Lambda) = \dfrac{T_{\rm obs}}{N_{\rm gen}} \sum_{j=1}^{N_{\rm thr}}s_j\,;
    \label{eq:approx_N_det}
\end{equation}
where $N_{\rm gen}$ is the total number of generated injections, $N_{\rm thr}$ is the number of injections with $\rho>\rho_{\rm thr}$, and $s_j$ denotes the importance-sampling weight of the j-th detected injection.

The injection $\pi_{\rm inj}$ can be chosen so as to simplify the expression in Eq.~(\ref{eq:approx_N_det}). In particular, a convenient choice is to use the normalized merger rate itself as injection prior. In this case, we define
\begin{equation}
    \pi_{\rm inj} = \dfrac{T_{\rm obs}}{N_{\rm id}}\dfrac{\text{d}N}{\text{d}t_{\rm d}\text{d}\theta}\Rightarrow s_j = \dfrac{N_{\rm id}}{T_{\rm obs}}\,.
\end{equation}
With this specific definition of $\pi_{\rm inj}$, the estimator in Eq.~(\ref{eq:approx_N_det}) simplifies to
\begin{equation}
    N_{\rm det}\approx \dfrac{1}{N_{\rm gen}} \sum_{j=1}^{N_{\rm thr}} N_{\rm id} = N_{\rm id} \dfrac{N_{\rm thr}}{N_{\rm gen}} \,.
    \label{eq:approx_N_det2}
\end{equation}
Introducing the ratio $r = N_{\rm thr}/N_{\rm gen}$, the expected number of detected events can therefore be written as $N_{\rm det}=r\,N_{\rm id}$.
To determine a sufficient number of injections $N_{\rm gen}$, we follow the effective-sample criterion introduced in \cite{Farr:2019Neff_inj}. The \emph{effective number of injections} $N_{\rm eff,inj}$ is defined as
\begin{equation}
    N_{\rm eff,inj} = \dfrac{\left(\sum_{j}^{N_{\rm thr}} s_j\right)^2}{\sum_{j}^{N_{\rm thr}} s_j^2 - N_{\rm gen}^{-1}\left(\sum_{j}^{N_{\rm thr}} s_j\right)^2} = N_{\rm gen}\dfrac{N_{\rm thr}/N_{\rm gen}}{1 - N_{\rm thr}/N_{\rm gen}}=N_{\rm gen}\dfrac{r}{1-r}\,.
    \label{eq:N_eff_inj}
\end{equation}

Following \cite{Farr:2019Neff_inj}, we require $N_{\rm eff,inj} > 4 N_{\rm obs}$. We generate $N_{\rm gen}=4N_{\rm id}$ injections for each $\Lambda$ (see Appendix~\ref{app:N_gen}), so that $N_{\rm eff,inj}=4N_{\rm id}r/(1-r)\approx6N_{\rm id}$ for $r\approx0.6$ (example value found from injections, cf. Tab. \ref{tab:N_res}), well above $4N_{\rm obs}\approx4rN_{\rm id}\approx 2.4 N_{\rm id}$ (assuming $N_{\rm obs}$ extracted from a Poisson distribution of parameter $N_{\rm det}>>1$ from Eq.~(\ref{eq:approx_N_det2})).

The Monte Carlo error for the $N_{\rm det}$ approximation in Eq.~(\ref{eq:approx_N_det}) is given by $\sigma_{\rm MC} = N_{\rm det}/\sqrt{N_{\rm eff,inj}}$ \cite{Farr:2019Neff_inj} and the ratio with the Poisson error for $N_{\rm obs}$ ($\sigma_{\rm P}=\sqrt{N_{\rm det}}$) gives $\sigma_{\rm MC}/\sigma_{\rm P} = \sqrt{N_{\rm det}/N_{\rm eff,inj}}$, which combined with our choice of $N_{\rm gen}$ gives: $\sigma_{\rm MC} = \sqrt{1-r}/2\approx 0.3$ Poisson widths, thus subdominant and confirming the choice of $N_{\rm gen}$. 

\section{Results}\label{sec:results}
In this section we apply the number-count analysis to a mock ET catalog (Sec.~\ref{sec:NC_PE}) and validate the inference with calibration tests on various mock datasets (Sec.~\ref{sec:mock_diag}).
\subsection{Number counts parameter estimation}\label{sec:NC_PE}

\begin{figure}[h]
    \centering
    \includegraphics[width=0.6\textwidth]{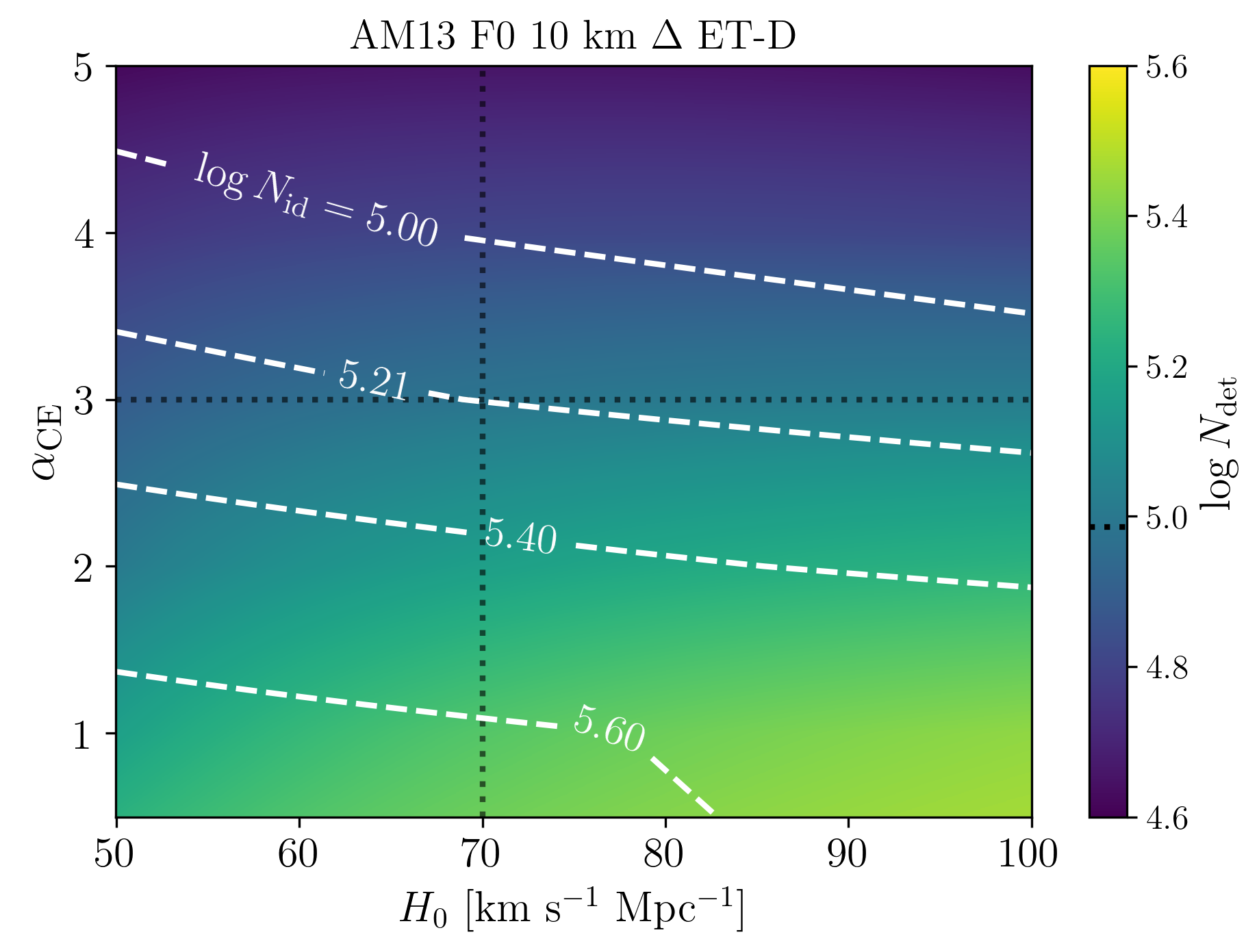}
    \caption{Logarithm of the expected number of detectable events $\log N_{\rm det}$ in a single year, as a function of the Hubble constant $H_0$ and the common-envelope efficiency $\alpha_{\rm CE}$, for the ET-D sensitivity curve of the triangular $(\Delta)$ configuration and the fiducial merger rate density model (AM13 F0). The color map represents $\log N_{\rm det}$, while the white dashed lines show iso-contours of constant $N_{\rm id}$ (total number of events for an ideal detector). The black dotted lines mark the fiducial parameter values $\Lambda_{\rm F}=(H_0^{\rm F},\alpha_{\rm CE}^{\rm F})=(70\text{ km s$^{-1}$Mpc$^{-1}$}, 3)$ for which $\log N_{\rm det}^{\rm F} = 4.99$. The color bar shows the corresponding values of $\log N_{\rm det}$.}
    \label{fig:AM13_N_det} 
\end{figure}

To obtain $N_{\rm det}(\Lambda)$, we first computed $N_{\rm det}(H_0)$ through different injection campaigns varying the generated samples according to $N_{\rm gen} = 4 N_{\rm id}(H_0,\alpha_{\rm CE})$ (Sec.~\ref{sec:BBH_NC}) for each simulated $\alpha_{\rm CE}$ and 10 values of $H_0$ $\in[50, 100]$ km s$^{-1}$Mpc$^{-1}$ on a linear grid, then we took a linear fit of $\log N_{\rm det}(H_0)$ (at fixed common-envelope efficiency) and finally, a 2D linear interpolation was constructed between the different $\alpha_{\rm CE}$ curves to speed up the analysis. With this construction, the function $N_{\rm det}(\Lambda)$ is shown in Fig. \ref{fig:AM13_N_det} for the fiducial AM13 F0 model. Overlaid we also plotted curves of constant $N_{\rm id}$ as white dashed lines. The same procedure to create $N_{\rm det}(\Lambda)$ for the fiducial model is also applied to all the additional models discussed in Subsec.~\ref{sec:MRD_detail} and we use them in the analysis. In Tab.~\ref{tab:N_res} we give our estimates for the total number of events per year, those actually expected and the fraction of detectable events at fiducial parameters for ET-D $\Delta$ 10 km.  

For our mock observation of dark sirens, we adopt a set of fiducial parameter values $\Lambda_{\rm F}$, corresponding to the fiducial cosmology introduced in Sec.~\ref{sec:SFRD} and $\alpha_{\rm CE}^{\rm F}=3$. CE evolution is still poorly constrained. Our fiducial value is supported by population-synthesis comparisons with GW data \cite{Santoliquido:2021MRDs}, by X-ray binaries \cite{Kalogera:1999alphaXrB1, Podsiadlowski:2003alphaXrB2, Li:2026Xray_alpha} and by studies of CE energetics \cite{Ivanova:2013CE, Ivanova:2017CE_m1}. While this fiducial model overpredicts the GWTC-5.0 local merger rate by a factor of a few (see Sec. \ref{sec:pop_synth_uncertain}), this does not affect the internal consistency of the forecast, because the same model generates and analyzes the mock data. Using these fiducial parameters we find a value of $N_{\rm det}^{\rm F} = 96698 \approx 10^{4.99}$ merger events in a year (also shown in Fig.~\ref{fig:AM13_N_det} for the intersection of black dotted lines), from which we generate a mock observed number of events $N_{\rm obs}$ extracted from $\mathcal{P}oisson(N_{\rm det}^{\rm F})$. 

We evaluate the number likelihood, Eq.~(\ref{eq:number_like}), on a grid in $(H_0,\alpha_{\rm CE})$ with flat priors: $H_0\in[60,80]$\,km\,s$^{-1}$\,Mpc$^{-1}$, which brackets current measurements \cite{Planck:2016cosmo_pars, Riess:2021SH0ES}, and $\alpha_{\rm CE}\in[0.5,5]$, the range covered by our simulations. We analyze the same mock $N_{\rm obs}$, drawn from the fiducial AM13 F0 model, with all six models. Since the likelihood depends only on $N_{\rm det}(\Lambda)$, we expect a narrow degeneracy curve along $N_{\rm det}(\Lambda)=N_{\rm obs}$, with a relative width $1/\sqrt{N_{\rm obs}}\approx0.32\%$.

\begin{figure}[h]
    \centering
    \includegraphics[width=0.8\textwidth]{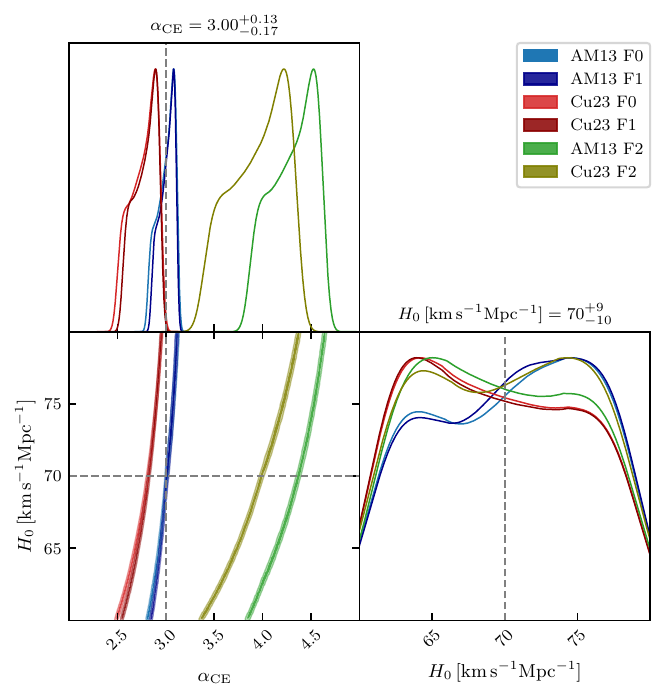}
    \caption{ Number-counts posteriors for initial flat priors $H_0 \in [60, 80]$\,km\,s$^{-1}$\,Mpc$^{-1}$ and $\alpha_{\rm CE}\in[0.5, 5]$. The number of observed events $N_{\rm obs}$ is extracted from a Poisson distribution with the AM13 F0 model and fiducial parameters, while inference is performed using the AM13-Cu23 F0, 1, 2 models. Single parameter estimates refer to median and $2\sigma$ confidence intervals for the fiducial model only.}    
    \label{fig:2d_flat} 
\end{figure}

In Fig. \ref{fig:2d_flat} we present the contribution to the hierarchical posterior coming from number counts and plot the 2D contours of the $1$ and $2\,\sigma$ credible regions and their marginal distributions for different FMR-factorization models. The posterior on $H_0$ follows the flat prior, which means that counts alone do not constrain $H_0$. In contrast, the constraint on $\alpha_{\rm CE}$ is very tight. For the AM13 F0 model we find, at $2\sigma$, $\alpha_{\rm CE}=3.00^{+0.13}_{-0.17}$.

It is worth noting that counts measure a single number, so $\alpha_{\rm CE}$ is constrained only through the overall normalization of the merger rate. Any other ingredient that changes this normalization, such as the binary fraction, the IMF, the SFR normalization, natal kicks or the stability of mass transfer, may be degenerate with $\alpha_{\rm CE}$. Moreover, the width of the $\alpha_{\rm CE}$ posterior is set by the projection of the $H_0$ prior along the degeneracy curve, not by the Poisson noise. Our constraints are therefore the statistical floor of number counts when all other ingredients are known.

Because the SFR density is anchored to galaxy counts, its $H_0^3$ scaling cancels the $H_0^{-3}$ scaling of ${\rm d}V_c/{\rm d}z$, so $N_{\rm id}$ depends on $H_0$ only through the age-redshift relation. The residual dependence of $N_{\rm det}$ on $H_0$ comes from the selection function, through $d_L\propto H_0^{-1}$.

Only AM13 F0 and F1 recover the true values within $2\sigma$, while the other models are offset. To quantify the offset of each model, we measure the distance of the true parameters from the posterior ridge in units of its width. To first order this is $\delta_M=|N^M_{\rm det}(\Lambda_{\rm F})-N_{\rm obs}|/\sqrt{N_{\rm obs}}$, reported in Tab.~\ref{tab:N_res}. We find offsets from $O(10)$ (Cu23-F0,1) to $O(100)$ (AM13/Cu23-F2) of Poisson widths, dominating the Poisson fluctuation, highlighting the importance of assessing the correct astrophysical model before any cosmo-astrophysical inference. 

\begin{figure}[h]
    \centering
    \includegraphics[width=0.8\textwidth]{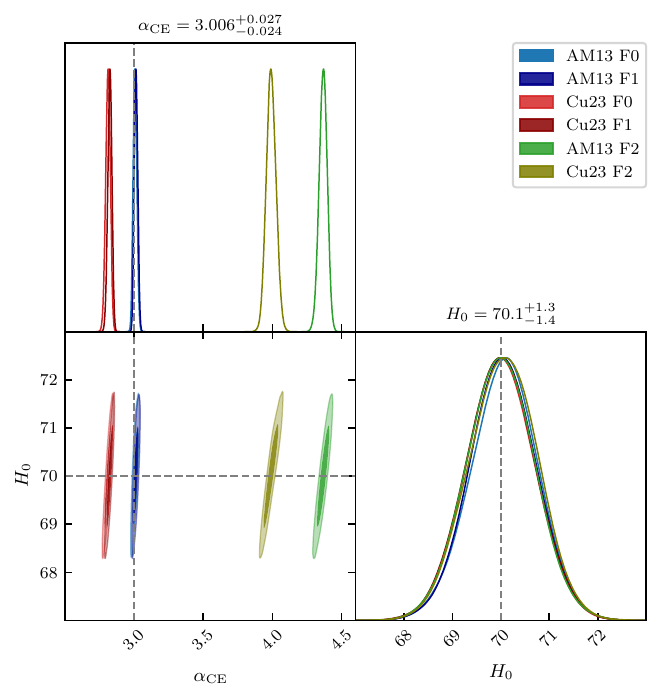}
    \caption{ Posterior coming from number counts for a $1\%$ $H_0$ Gaussian prior and flat prior $\alpha_{\rm CE}\in[0.5, 5]$. The number of observed events $N_{\rm obs}$ is extracted from a Poisson distribution with the AM13 F0 model and fiducial parameters, while inference is performed using the AM13-Cu23 F0,1,2 models. Single parameter estimates refer to median and $2\sigma$ confidence intervals for the fiducial model only.}
    \label{fig:2d_gaus} 
\end{figure}

Given the tightness of the degeneracy, in Fig. \ref{fig:2d_gaus} we test a $1\%$ Gaussian prior on $H_0$ and its effect on the posterior of $\alpha_{\rm CE}$. When the correct astrophysical model is considered, we find $\alpha_{\rm CE}=3.006^{+0.027}_{-0.024}$ ($2\sigma$). This sub-percent measurement shows the combined power of forward modeling, the large ET population, and precision cosmology. Finally, in all cases we notice that for both FMRs the F0 and F1 factorizations are degenerate, and in Sec. \ref{sec:mock_diag} we further analyze whether this degeneracy is due to the choice of fiducial values or a lack of correlations between mass variables and delay times.

\begin{table}[htbp]
\centering
\begin{tabular}{|c|c|c|c|c|}
\hline
\textbf{Model}& $\log N_{\rm id}(\Lambda_{\rm F})$ & $\log N_{\rm det}(\Lambda_{\rm F})$ & $r(\Lambda_{\rm F})$ & $\delta_M$\\
\hline
AM13 F0 & 5.21 & 4.99 & 60 \% & 0.73\\
AM13 F1 & 5.21 & 4.99 & 60 \% & 1.21\\
AM13 F2 & 5.46 & 5.13 & 47 \% & 106.19\\
Cu23 F0 & 5.21 & 4.95 & 55 \% & 25.06\\ 
Cu23 F1 & 5.21 & 4.95 & 55 \% & 25.07\\
Cu23 F2 & 5.45 & 5.09 & 44 \% & 84.22\\
\hline
\end{tabular}
\caption{Single year forecast values for ET $\Delta$ 10 km at fiducial values $\Lambda_{\rm F}$: total number of BBH events $N_{\rm id}$, number of detected events $N_{\rm det}$, fraction of detectable events $r$ and $\delta_M$ the offset of the true parameters from the posterior ridge, in Poisson widths.}
\label{tab:N_res}
\end{table}

\subsection{Calibration tests}\label{sec:mock_diag}

\begin{figure}[h]
    \centering
    \includegraphics[width=0.8\textwidth]{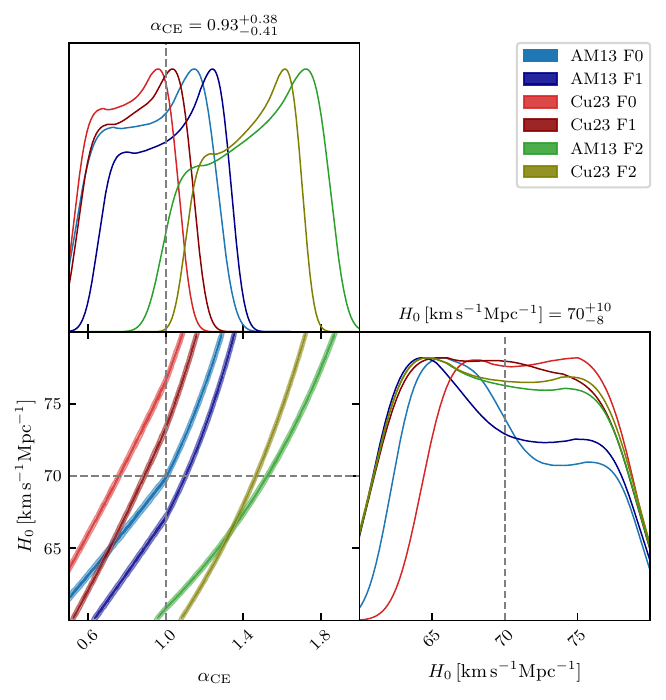}
    \caption{Posteriors coming from number counts for initial flat priors $H_0 \in[60, 80]$\,km\,s$^{-1}$\,Mpc$^{-1}$ and $\alpha_{\rm CE}\in[0.5, 5]$. The number of observed events $N_{\rm obs}$ is extracted from a Poisson distribution with the AM13 F0 model with $\alpha_{\rm CE}=1$ as the fiducial astrophysical parameter, different with respect to Fig. \ref{fig:2d_flat}; the resulting posteriors are all distinct. Single parameter estimates refer to median and $2\sigma$ confidence intervals for the fiducial model only.}
    \label{fig:2d_flat_aCE1} 
\end{figure}

For both FMRs, F0 and F1 give overlapping posteriors in Fig.~\ref{fig:2d_flat}. One might conclude that correlations between masses and delay times are negligible. This is not general because if we generate the mock data with $\alpha_{\rm CE}=1$, the posteriors of all models are separate (Fig.~\ref{fig:2d_flat_aCE1}). For this fiducial value, a flat prior on $H_0$ gives $\alpha_{\rm CE}=0.93^{+0.38}_{-0.41}$ ($2\sigma$, Fig. \ref{fig:2d_flat_aCE1}), i.e. $\approx40\%$ error, whereas a 1\% Gaussian prior on $H_0$ was found to tighten the constraint up to $\simeq 7\%$ ($2\sigma$) error on $\alpha_{\rm CE}$. To test miscalibration independently of the fiducial values, we use the diagnostics described below. We call model $M$ a combination of FMR and factorization, and $M_0=$ AM13 F0 the model that generates the mock data.

\subsubsection{Miscalibration}

Miscalibration can induce biases in posterior distributions whenever the assumed model differs from the true underlying model that explains the data, which in a mock inference is the one used to extract the mock dataset. We test it with a simulation-based calibration and construct probability-probability plots comparing the distribution of credible regions for the FMR-factorization model $M$ (with $M_0=$ AM13 F0), namely $\beta^M$s, with their expected calibrated distribution $\mathcal{U}(0,1)$. The method, based on the probability integral transform, is described in App.~\ref{sec:pp_plots}.

In Tab. \ref{tab:beta_KS} we present the value of the Kolmogorov-Smirnov statistic (maximum distance between the cumulative distribution function of the $\beta^M$s and the one from the uniform distributions). We clearly find that any model different from $M_0$ produces a statistically significant miscalibration, with negligible $p$-values for all models but AM13 F0, for which $p=0.85$. The miscalibration is larger for the wrong FMR and, for a fixed FMR, increases from F0 to F2.

\begin{table}[htbp]
  \centering
  \begin{tabular}{|c|c|c|c|}
  \hline
  \textbf{FMR}& F0 & F1 & F2 \\
    \hline
    AM13 & 0.019 & 0.851 & 0.994\\
    Cu23 & 0.928 & 0.994 & 0.999\\
    \hline
  \end{tabular}
  \caption{Measure of miscalibration with KS-statistic values for different factorizations and FMR models.} 
  \label{tab:beta_KS}
\end{table}

The joint test is far more sensitive than the marginal ones discussed below. Since the posterior is a thin curve, an offset of about one Poisson width across it moves the true parameters outside the high-density region, while it barely shifts the marginals, whose width is set by the extent of the curve within the prior.

\subsubsection{Bias}
From the KS test on the joint two dimensional posterior we find whether a model is generally miscalibrated. The directionality of the miscalibration is instead encoded in the one dimensional marginal posteriors where there is a proper ordering definition. Using again the samples from the prior $\Lambda_k$, we test for biases as follows:
\begin{itemize}
    \item[1)] We compute the marginal posteriors
    \begin{equation}
        P^M_k(\Lambda^{(i)}) = \int \text{d}\Lambda^{(j\neq i)} P_k^M(\Lambda)\,,(\Lambda^{(1)}= \alpha_{\rm CE}, \Lambda^{(2)}= H_0)\,.
    \end{equation}
    \item[2)] Obtain the cumulative distribution functions $F_{\Lambda^{(i)}}^M(t)$, and the credible levels
    \begin{equation}
        p_k^{M, (i)} = F_{\Lambda^{(i)}}^M(\Lambda_k^{(i)}) = \int^{\Lambda_k^{(i)}}_{\Lambda^{(i)}_{\rm min}} \text{d}\Lambda^{(i)} P^M_k(\Lambda^{(i)})\,,
    \end{equation}
    with $\Lambda^{(i)}_{\rm min}$ the lower bound of the $\Lambda^{(i)}$ prior. Since $p_k^{M,(i)}\sim \mathcal{U}(0,1)$ when $M=M_0$ (analogous to the $\beta^M$ case), we have that $w_k^{(i)} = \Phi^{-1}(p_k^{M_0,(i)})\sim\mathcal{N}(0,1)$, where $\Phi^{-1}(y)$ is the inverse normal cumulative distribution function. With this method we can identify the direction of bias and the distance from the truth for miscalibrated models with deviations from the normal distribution.
\end{itemize}

\begin{figure}[ht]
    \centering
    \includegraphics[width=1\textwidth]{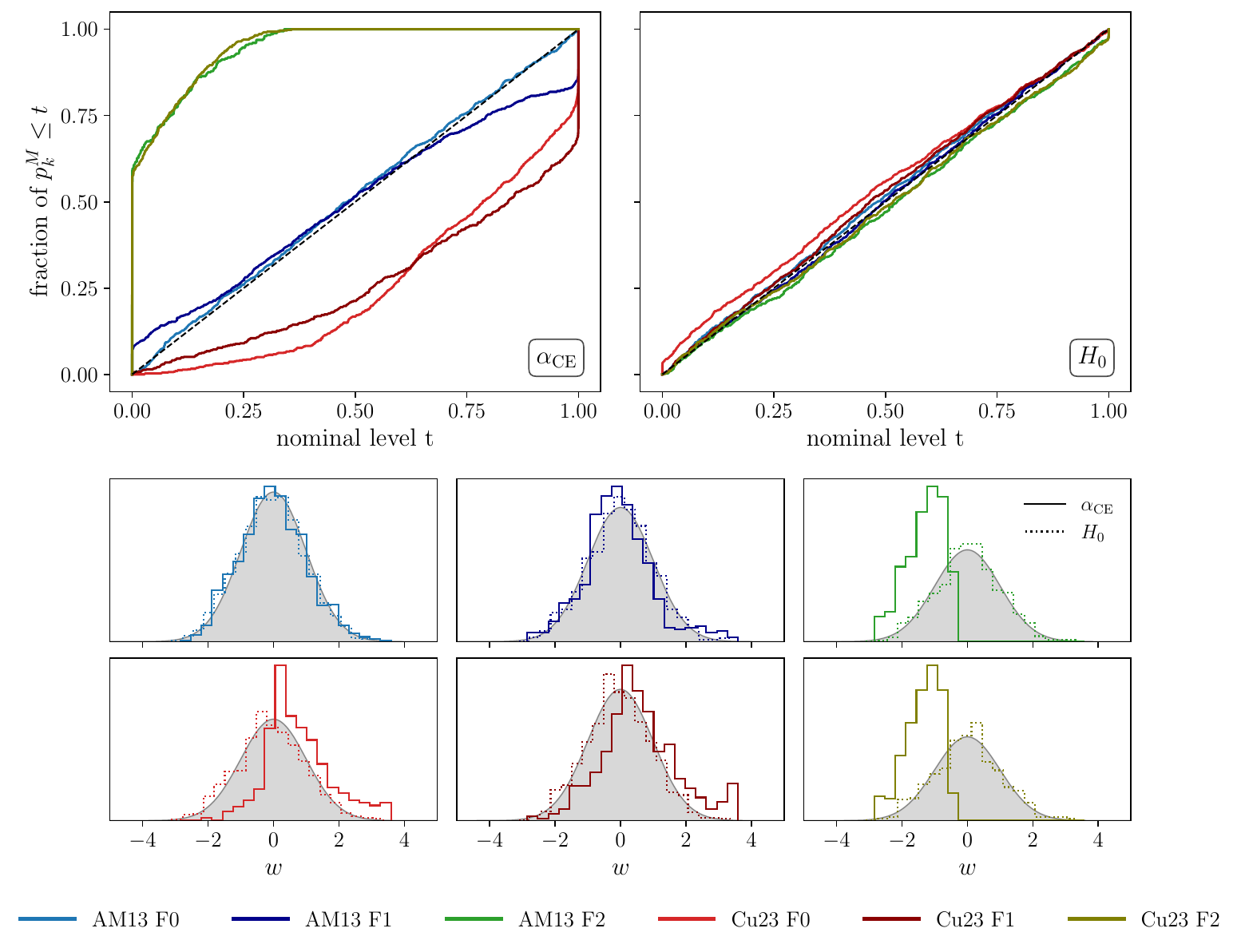}
    \caption{Qualitative bias validation tests on the marginal posterior distributions of every FMR-factorization model. \emph{Top panel}: comparison of the empirical cumulative distribution function of the credible levels $p_k$ with the uniform CDF. \emph{Bottom panel}: comparison of the histogram for the transformed credible levels $w_k$ with the normal distribution $\mathcal{N}(0,1)$.}
    \label{fig:bias_panel} 
\end{figure}

In Fig. \ref{fig:bias_panel} in the top panel we plot the empirical cumulative distribution function (CDF) of the credible levels $p_k^M$ of the marginal posteriors and compare them with the diagonal CDF of a uniform distribution. In the bottom panel we show the histograms of the transformed credible levels and compare them to the normal distribution to identify the direction of the bias. In Tab. \ref{tab:bias_table} we give for each model the respective Kolmogorov-Smirnov distance with respect to the ideal distribution (either uniform for $p_k$ or normal for $w$, having the same KS distance due to the monotonicity of $\Phi^{-1}(y)$), the median and the standard deviation of the $w$s. 

Two distinctive pictures appear for the population parameters considered when using only number-count posteriors. The common-envelope efficiency shows higher sensitivity to miscalibration irrespective of the initial fiducial values and the $p$-value from the KS distance is negligible for all models but  AM13 F0, for which $p=0.48$. On the contrary, $H_0$ is insensitive to the underlying astrophysics and no model is disfavored, a direct consequence of the prior-dominated $H_0$ posterior. For the same reason, biases can be measured reliably only for $\alpha_{\rm CE}$. We compute the bias as ${\rm median}(w)/\sigma_w$, a dimensionless number expressed in units of $\sigma_w$: a negative bias means that the truth tends to lie in the lower tail of the posterior, i.e. the posterior overestimates the parameter. 

The AM13 F0 and F1 models are very similar: their biases on $\alpha_{\rm CE}$ are both small ($-0.04$ and $-0.08$, respectively), far below the width of the posterior, which explains the overlapping posteriors in Fig.~\ref{fig:2d_flat}. The two models are nevertheless not identical: the KS test rejects AM13 F1 ($p$-value $=2.9\times10^{-3}$), as its $w$ distribution is slightly broader than a unit normal ($\sigma_w=1.21$ against $1.01$). F2 is strongly negatively biased for both FMRs ($-1.62$ and $-1.95$), and pushes the posteriors towards the upper edge of the $\alpha_{\rm CE}$ prior. Cu23 F0 and Cu23 F1 have large positive biases ($0.55$ and $0.37$), i.e. they tend to underestimate $\alpha_{\rm CE}$. The bias of Cu23 F1 is smaller than that of Cu23 F0 because neglecting the mass--delay-time correlations partly compensates for the wrong FMR, so that a model with two wrong ingredients can mimic a more correct one (see also Fig.~\ref{fig:2d_flat_aCE1}).

\begin{table}[ht]
\centering
\begin{tabular}{|llcccc|}
\hline
\textbf{Parameter} & \textbf{Model} & KS distance & median $w$ & $\sigma_w$ & bias\\
\hline
$\alpha_{\rm CE}$ & AM13 F0 & 0.027 & $-0.04$ & 1.01 & -0.04\\
                  & AM13 F1 & 0.068 & $-0.10$ & 1.21 & -0.08\\
                  & AM13 F2 & 0.661 & $-1.26$ & 0.78 & -1.62\\
                  & Cu23 F0 & 0.314 & $0.61$  & 1.10 & 0.55\\
                  & Cu23 F1 & 0.222 & $0.48$  & 1.30 & 0.37\\
                  & Cu23 F2 & 0.686 & $-1.29$ & 0.66 & -1.95\\
\hline
$H_0$             & AM13 F0 & 0.026 & $-0.05$ & 1.01 & -0.05\\
                  & AM13 F1 & 0.018 & $-0.01$ & 1.00 & -0.01\\
                  & AM13 F2 & 0.035 & $0.04$  & 1.01 & 0.04\\
                  & Cu23 F0 & 0.053 & $-0.11$ & 1.08 & -0.1\\
                  & Cu23 F1 & 0.039 & $-0.09$ & 1.00 & -0.09\\
                  & Cu23 F2 & 0.023 & $0.02$  & 1.05 & 0.02\\
\hline
\end{tabular}
\caption{Quantitative bias test on the marginal posterior distributions for every FMR-factorization model. For both parameters we compute: i) the Kolmogorov-Smirnov distance of the $p_k^M$ empirical CDF from the uniform distribution (same values for the KS distance of the $w_k$ with respect to the normal distribution due to monotonicity of $\Phi^{-1}(y)$); ii) median on $w$ and standard deviation $\sigma_{w}$ of the transformed credible levels $w$; iii) the bias, defined as $\mathrm{median}(w)/\sigma_w$ (dimensionless, in units of $\sigma_w$).}
\label{tab:bias_table}
\end{table}

\section{Comparison with GW data}\label{sec:discussion}
In this section we compare our models with the GWTC-5.0 local merger rate. We ask whether the excess we find comes from the star formation history (Sec.~\ref{sec:sfrd_alt}) or from binary evolution (Sec.~\ref{sec:pop_synth_uncertain}).

\subsection{SFR density alternative model}\label{sec:sfrd_alt}
In Subsec.~\ref{sec:SFRD} we presented our data-driven framework for the SFR density. A commonly used alternative in the literature is the analytic fit proposed by Madau and Fragos \cite{Madau:2017_SFRD},
\begin{equation}
    \rho_{\rm SFR}^{\rm MF17}(z) = 0.01 \dfrac{(1+z)^{2.6}}{1+[(1+z)/3.2]^{6.2}}\,.
    \label{eq:MF_SFRD}
\end{equation}

Since this model is independent of metallicity, it must be supplemented with a distribution in $Z$ in order to interface it with binary population synthesis simulations (from Subsec.~\ref{sec:simulations}). Following Ref. \cite{Sgalletta:2025high_MRD}, we assume a log-normal distribution centered on the average metallicity of the universe at redshift $z$, namely $\Braket{\log(Z/Z_{\odot})}$ :

\begin{equation}
    \dfrac{\text{d}p}{\text{d}\log Z}(Z|z) = \dfrac{1}{\sqrt{2\pi\tilde{\sigma}_Z^2}}e^{-\dfrac{[\log (Z/Z_{\odot}) - \Braket{\log(Z/Z_{\odot})} ]^2}{2\tilde{\sigma}_Z^2}}\,,
\end{equation}
which leads to the SFR density per unit metallicity
\begin{equation}
    \dfrac{\mathrm{d}\rho_{\rm SFR}^{\rm MF17}}{\mathrm{d}\log Z}(Z, z) = \rho_{\rm SFR}^{\rm MF17}(z) \dfrac{\text{d}p}{\text{d}\log Z}(Z|z)\,,
    \label{eq:dSFRD_dlogZ_MF}
\end{equation} 
In Eq.~(\ref{eq:dSFRD_dlogZ_MF}) the mean metallicity evolves with redshift as
\begin{equation}
    \Braket{\log(Z/Z_{\odot})} = 0.153 - 0.074z^{1.34} -\ln(10) \dfrac{\tilde{\sigma}_Z^2}{2}\,,
\end{equation}
and $\tilde{\sigma}_Z$ represents the intrinsic scatter of this metallicity distribution, which is distinct from the dispersion $\sigma_{\rm FMR}$ used in our data-driven framework for the SFR density (Subsec.~\ref{subsec:SFRD_model}). 

We compute the merger rate with this prescription through Eqs.~(\ref{eq:diff_MRD})--(\ref{eq:MRD}) and compare it with the BBH merger rate at redshift $z=0.2$ from the Gravitational-Wave Transient Catalog 5.0 (GWTC-5.0) \cite{LVK:2026GWTC5_intro, LVK:2026GWTC5_obs, LVK:2026GWTC5_pop}, $\mathcal{R}_{0.2}^{\rm LVK} = 30.4^{+6.5}_{-5.4}$ Gpc$^{-3}$\,yr$^{-1}$, obtained with the \textsc{FullPop} model \cite{LVKAbac:2025GWTC4cosmo, Fishbach:2020Fullpop_1, Farah:2022Fullpop_2, Mali:2025Fullpop_3} a minimum BH mass of $3 M_{\odot}$ \cite{LVK:2026GWTC5_pop}, as in our simulations, and a maximum BH mass of $300 M_{\odot}$ defined in Ref. \cite{LVK:2026GWTC5_pop}. For consistency with that analysis, we compute our rates with the Planck15 cosmological parameters ($H_0=67.9$\,km\,s$^{-1}$\,Mpc$^{-1}$, $\Omega_{\rm m,0}=0.3065$) \cite{Planck:2016cosmo_pars} and denote them $\mathcal{R}_{0.2,\rm Pl15}$.

\begin{figure}[h]
    \centering
    \includegraphics[width=0.6\textwidth]{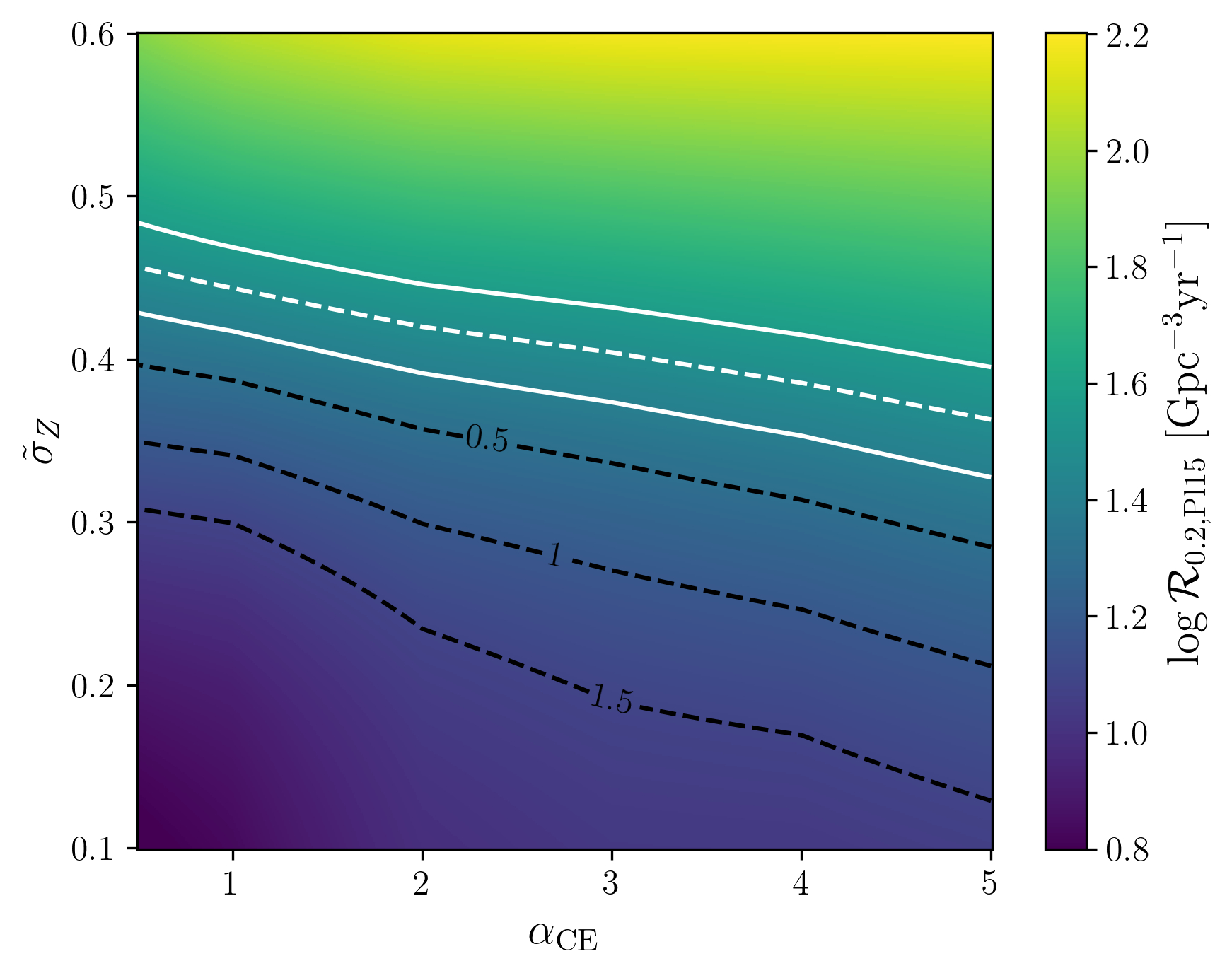}
    \caption{ Logarithm of the predicted merger rate density at $z=0.2$ with Planck15 cosmology \cite{Planck:2016cosmo_pars}, $\log \mathcal{R}_{0.2,\rm Pl15}$, as a function of the common-envelope efficiency $\alpha_{\rm CE}$ and metallicity scatter $\tilde{\sigma}_Z$, for the Madau-Fragos SFR density model in Eq.~(\ref{eq:dSFRD_dlogZ_MF}). The color map shows the value of $\log \mathcal{R}_{0.2,\rm Pl15}$. The white dashed contour corresponds to $\mathcal{R}_{0.2,\rm Pl15} = \mathcal{R}_{0.2}^{\rm LVK}$, while the white solid contours indicate the $90\%$ credible intervals. Black dashed contours represent the rescaled relation $\mathcal{R}_{0.2,\rm Pl15} = \bar{\mathcal{R}}_{0.2,\rm Pl15}(1+\delta \mathcal{R}_{0.2, \rm Pl15})$ for $\delta \mathcal{R}_{0.2, \rm Pl15} = 0.5,1,1.5$, illustrating the likely effect of additional formation channels.}
    \label{fig:R0_MD}
\end{figure}

Figure~\ref{fig:R0_MD} shows $\log\mathcal{R}_{0.2,\rm Pl15}$ for the MF17 model in the $(\alpha_{\rm CE},\tilde\sigma_Z)$ plane. The white dashed curve marks $\mathcal{R}_{0.2,\rm Pl15}=\mathcal{R}^{\rm LVK}_{0.2}$ and the solid curves its 90\% credible interval. The rate changes by a factor of about 2 across the $\alpha_{\rm CE}$ range (see also Fig.~\ref{fig:MRD_comp}), but by more than an order of magnitude with $\tilde\sigma_Z$. Smaller scatters give more weight to metal-rich populations, which produce fewer mergers within a Hubble time, as also noted by \cite{Sgalletta:2025high_MRD}.

For all simulated $\alpha_{\rm CE}$, reproducing the observed rate requires $\tilde{\sigma}_Z<0.5$. This holds if all mergers originate from isolated binaries. If additional channels contribute, the predicted local merger rate could be rewritten as $\mathcal{R}_{0.2,\rm Pl15} = \bar{\mathcal{R}}_{0.2,\rm Pl15} (1 + \delta \mathcal{R}_{0.2, \rm Pl15})$, where $\bar{\mathcal{R}}_{0.2,\rm Pl15}$ is the merger rate computed from our isolated binary simulations and $\delta \mathcal{R}_{0.2,\rm Pl15}$ quantifies the fractional contribution from other channels. As shown by the black dashed contours in Fig.~\ref{fig:R0_MD}, including such contributions further reduces the allowed values of $\tilde{\sigma}_Z$ to the region below $0.5$. This requirement is in tension with independent observational constraints, which suggest a broader metallicity distribution with $\tilde{\sigma}_Z> 0.5$ \cite{Chruslinksa:2024Z_avg, Sgalletta:2025high_MRD}. Within our isolated-binary framework, the combination of the Madau-Fragos SFR density with a log-normal metallicity distribution therefore cannot reproduce the local merger rate and the observed metallicity spread at the same time. Since the predicted rate depends on the full combination of SFR, metallicity distribution, merger efficiency and delay times, this tension cannot be attributed to the SFR density alone.

We also compared the redshift distribution of mergers predicted by our fiducial model with that of MF17, using the Kullback-Leibler divergence \cite{Kullback:1951KL_div} between the normalized distributions. The divergence decreases with $\tilde\sigma_Z$ up to the edge of the explored range ($\tilde\sigma_Z=0.6$, set by our metallicity grid): in shape, our data-driven framework behaves as MF17 with a broad metallicity spread, as expected from its construction on galaxy scaling relations.

Thus, neither MF17 nor our data-driven SFR density reproduces the GWTC-5.0 local rate with a metallicity spread consistent with galaxy observations ($\tilde\sigma_Z\gtrsim0.5$). The excess therefore cannot be attributed to the SFR density alone. In the next subsection we fix the SFR density to our framework and examine the role of binary evolution.

\subsection{Population synthesis uncertainties} \label{sec:pop_synth_uncertain}
We now fix the SFR density to our data-driven framework (Subsec. \ref{sec:SFRD}) and investigate the main sources of uncertainty arising from the isolated binary population synthesis model. 

A first indication of these uncertainties comes from the comparison between the predicted merger rate density of our fiducial model (AM13 F0), and the value inferred by the LVK collaboration. According to our fiducial model and our {\footnotesize\textup{\textrm{SEVN}}} simulations we obtain a merger rate density at $z\approx0.2$ for the Planck15 cosmology (Subsec. \ref{sec:sfrd_alt}): 
\begin{equation}
    \mathcal{R}_{0.2,\rm Pl15}\approx(278,227, 200, 158, 148, 122)\,\text{Gpc}^{-3}\text{yr}^{-1} \text{ for } \alpha_{\rm CE}\in[0.5,5]\,;
\end{equation}
these local merger rate density values are larger than $\mathcal{R}_{0.2}^{\rm LVK} = 30.4^{+6.5}_{-5.4}$ Gpc$^{-3}$yr$^{-1}$ by a factor of 4 to 9, depending on $\alpha_{\rm CE}$, and the discrepancy also persists with the Cu23 FMR. A similar trend was already reported in \cite{Sgalletta:2025high_MRD}, where different prescriptions for isolated binary evolution were found to increase the local merger rate density even further. 

In addition to these cases, we also reconsider the F2 model, which corresponds to a log-flat distribution for delay times. In this scenario, with the fiducial FMR model AM13 (Sec. \ref{sec:SFRD}), the prediction for the merger rate density is:
\begin{equation}
    \mathcal{R}_{0.2,\rm Pl15}\approx(202,164,113,80,61,47)\,\text{Gpc}^{-3}\text{yr}^{-1} \text{ for } \alpha_{\rm CE}\in[0.5,5]\,,
\end{equation}
and we notice that values $\alpha_{\rm CE}>5$ are still required to reproduce the inferred LVK rates. A log-linear fit of $\mathcal{R}_{0.2,\rm Pl15}(\alpha_{\rm CE})$, extrapolated to the GWTC-5.0 values, gives $\alpha_{\rm CE} \approx 13$ for F0 and $\approx 6$ for F2. These values lie far outside the simulated range. Moreover, $\alpha_{\rm CE}>1$ already requires energy sources other than the orbital energy, so such values mainly signal that the prescription itself may need to be revised. F2 gets closer to the data, but at the cost of many more mergers at high redshift, which only next-generation detectors can test.

Among the modeling choices considered here, the comparison between the factorizations indicates that the delay-time distribution has the largest impact on the predicted local rate. Indeed, the delay time strongly depends on the binary orbital separation, as wider systems lead to longer inspiral times. In isolated binary evolution, the orbital separation is primarily shaped by mass-transfer processes, and in particular by common-envelope evolution. To better understand this effect, we decompose our fiducial model into the main formation channels of BBHs in isolated binaries \cite{Broekgaarden:2022channels, Iorio:2023sevn_4, Sgalletta:2025high_MRD}:
\begin{itemize}
    \item Channel I: systems undergoing stable mass transfer before the formation of the first compact object, followed by at least one common-envelope phase;
    \item Channel II: systems evolving exclusively through stable mass transfer;
    \item Channel III: systems experiencing at least one common-envelope phase before the formation of the first compact object, involving a hydrogen-rich star and a stripped companion;
    \item Channel IV: similar to Channel III, but with both stars already stripped of their hydrogen envelopes.
\end{itemize}

We compare the merger rate density decomposition into these channels with the parametric model commonly used in GW data analysis, based on a Madau-like redshift evolution and a phenomenological mass distribution $p(m_1, m_2|\vec{\lambda})$:
\begin{equation*}
    \mathcal{R}(z|R_0,\gamma,\kappa, z_p, m_{\rm min}, m_{\rm max}) = R_0 [1 + (1+z_p)^{-\gamma-\kappa}]\dfrac{(1+z)^\gamma}{1+[(1+z)/(1+z_p)]^{\gamma+\kappa}}\times
\end{equation*}
\begin{equation}
    \times\int_{m_i\in[m_{\rm min}, m_{\rm max}]\,\wedge\,m_1\geq m_2} \text{d}m_1\text{d}m_2\,p(m_1,m_2|\vec{\lambda})\,.
    \label{eq:LVK_MRD}
\end{equation}

\begin{figure}[h]
    \centering
    \includegraphics[width=1.1\textwidth]{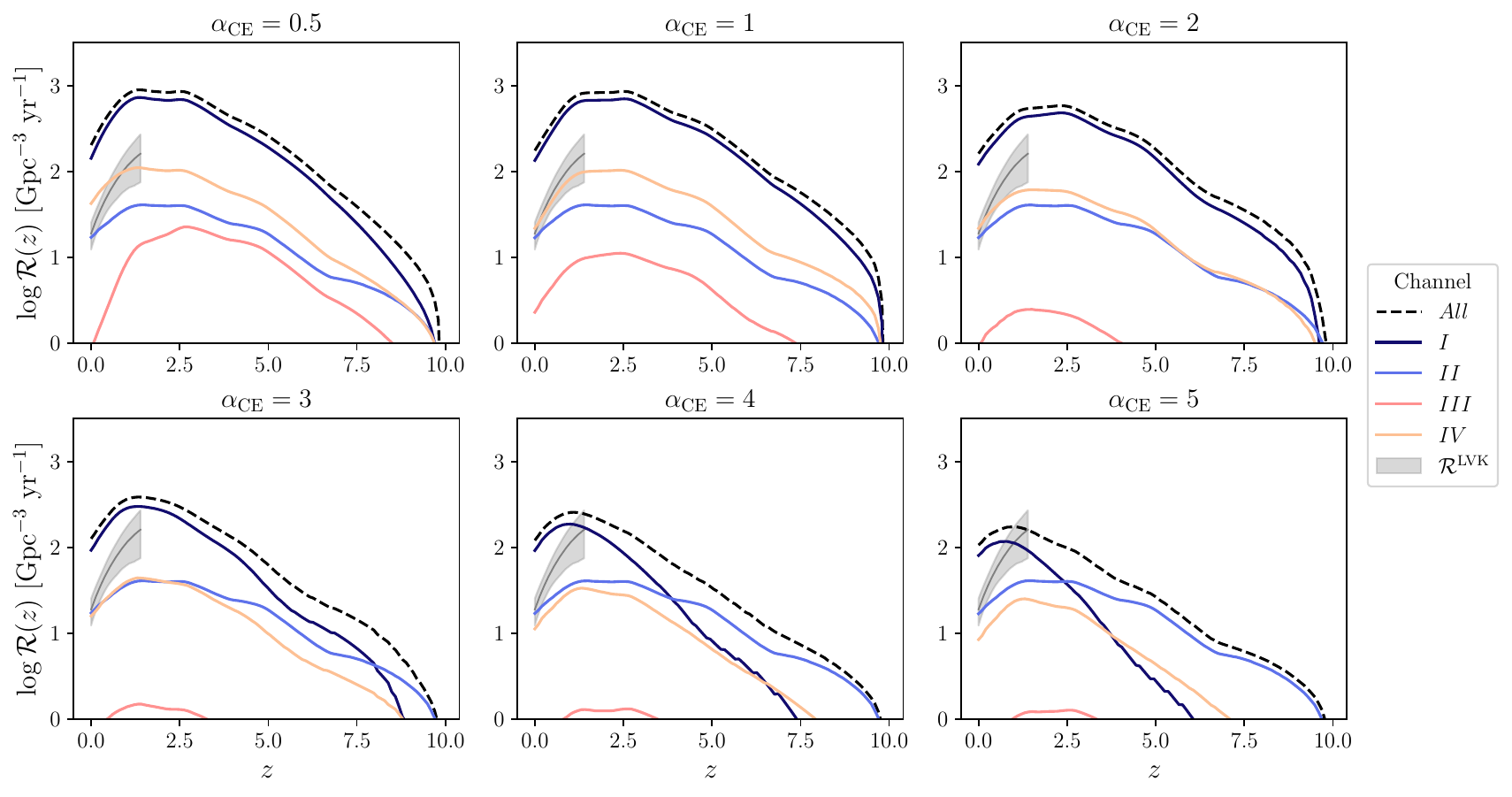}
    \caption{Semi-logarithmic plots of the merger rate density functions for different $\alpha_{\rm CE}$ with the fiducial model AM13 F0, assumed LVK cosmology (Subsec. \ref{sec:sfrd_alt}) and decomposed into the four isolated binary evolutionary channels. In gray color we present the median and $90\%$ curves inferred using the \textsc{FullPop} model \cite{LVKAbac:2025GWTC4cosmo, Fishbach:2020Fullpop_1, Farah:2022Fullpop_2, Mali:2025Fullpop_3} and the GWTC-5.0 catalog \cite{LVK:2026GWTC5_intro,LVK:2026GWTC5_obs, LVK:2026GWTC5_pop}, shown up to the redshift where $<1\%$ of BBH are found in the catalog \cite{LVK:2026GWTC5_pop} and after which the inferred model is less reliable.}
    \label{fig:MRD_channels}
\end{figure}

The parameters of this model are fitted to GWTC-5.0 observations \cite{LVK:2026GWTC5_intro, LVK:2026GWTC5_obs} using {\footnotesize\textup{\textrm{ICAROGW}}} \cite{Mastrogiovanni:2023icarogw, Mastrogiovanni:2023gal_cat} and the \textsc{FullPop} mass model \cite{LVKAbac:2025GWTC4cosmo, Fishbach:2020Fullpop_1, Farah:2022Fullpop_2, Mali:2025Fullpop_3}, over the mass range $m_i\in[3, 200]$ M$_{\odot}$\footnote{$R_0$ posterior samples have not been produced for the GWTC-5.0 data analysis (cf. fig. 5 from \cite{LVK:2026GWTC5_pop}). We therefore compute the rate curves with $R_0=1$ and normalized the resulting curves at $z=0.2$, drawing $\mathcal{R}_{0.2}$ from a split-normal distribution that matches $\mathcal{R}_{0.2}^{\rm LVK}= 30.4^{+6.5}_{-5.4}$ Gpc$^{-3}$yr$^{-1}$ \cite{LVK:2026GWTC5_pop}}.In Fig.~\ref{fig:MRD_channels} we show a comparison between: i) our fiducial model (AM13 F0) for the merger rate density curves with the Planck15 cosmology, decomposed in isolated formation channels at different $\alpha_{\rm CE}$ values and ii) the median and 90\% credible intervals of the merger rate density inferred from GWTC-5.0 data, up to the redshift $\hat{z}\approx1.4$ above which fewer than 1\% of the GWTC-5.0 BBHs lie $(z\approx 1.4)$ \cite{LVK:2026GWTC5_pop}.

\begin{figure}[h]
    \centering
    \includegraphics[width=1.1\textwidth]{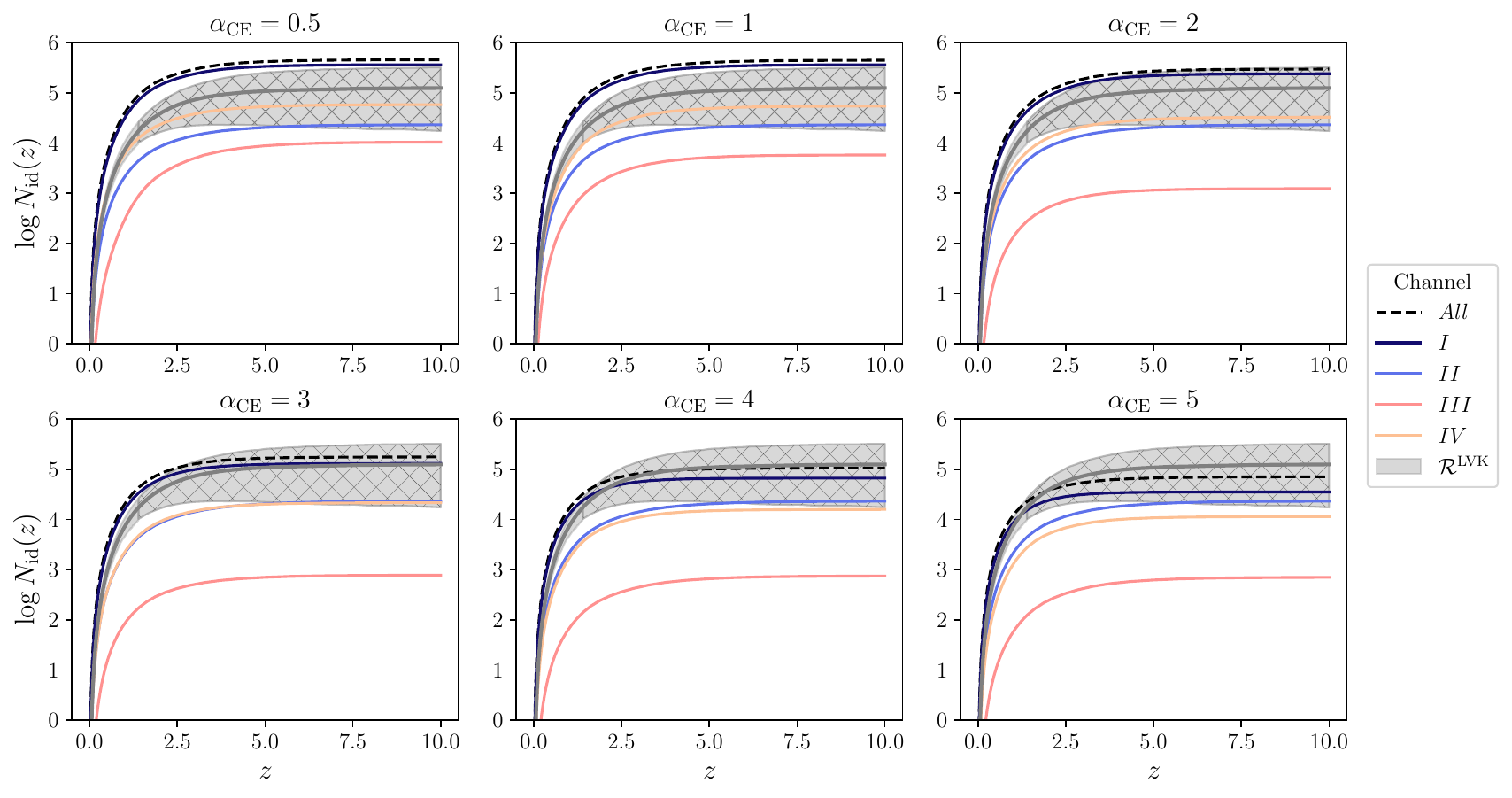}
    \caption{Semi-logarithmic plots of the cumulative number of total events compared for different $\alpha_{\rm CE}$ with the fiducial model AM13 F0, Planck15 cosmology, and decomposed into four different isolated binary evolution channels. In plain gray color we present the median and $90\%$ curves inferred using the \textsc{FullPop} model \cite{LVKAbac:2025GWTC4cosmo, Fishbach:2020Fullpop_1, Farah:2022Fullpop_2, Mali:2025Fullpop_3} and the GWTC-5.0 catalog \cite{LVK:2026GWTC5_intro,LVK:2026GWTC5_obs, LVK:2026GWTC5_pop} up to the redshift where $<1\%$ of BBH are found in the catalog \cite{LVK:2026GWTC5_pop} and in cross-hatched gray the model extrapolation above $z>\hat{z}$, where the inferred model is less reliable.}
    \label{fig:N_cum_ch}
\end{figure}

Additionally, in Fig.~\ref{fig:N_cum_ch} we decomposed in isolated formation channels the cumulative number of total events:
\begin{equation}
    N_{\rm id}(z|\mathbf{p}) = T_{\rm obs} \int_0^z \mathcal{R}(z'|\mathbf{p})\dfrac{\text{d}V_c}{\text{d}z'}\dfrac{1}{(1+z')}\text{d}z'\,,
\end{equation}
for a single year of observation, $\mathbf{p}$ as either $\alpha_{\rm CE}$ or the set of parameters in Eq.~(\ref{eq:LVK_MRD}) and we show the extrapolation of GWTC-5.0 after $\hat{z}$ as a cross-hatched region. 

Figure~\ref{fig:MRD_channels} shows that the excess in the local merger rate is mainly carried by Channel I, i.e. by systems undergoing stable mass transfer followed by a common-envelope phase. The cumulative number of events in fig.~\ref{fig:N_cum_ch} shows a better agreement for the $\alpha_{\rm CE} > 3$ models when all isolated formation channels are included, and these models would remain within the 90\% credible interval even if additional formation channels contributed. This agreement, however, is more evident at redshifts $z > \hat z$, where the GWTC-5.0 curve is an extrapolation of the parametric model of Eq.~(\ref{eq:LVK_MRD}) rather than a constraint from the data, and should therefore be regarded as indicative.

The dominance of Channel I in the local excess makes the common-envelope prescription a natural candidate for revision, though not the only one, since it has been shown that the local rate is also sensitive to: the stability criteria for mass transfer, the initial orbital-period distribution, natal kicks, as well as to the overall normalization set by the binary fraction and the IMF, and to the metallicity-dependent star formation history \cite{Sgalletta:2025high_MRD, Iorio:2023sevn_4, Broekgaarden:2022channels}. Exploring these ingredients is beyond the scope of this work. 

Overall, within the current $\alpha\lambda$ formalism, matching the observed rates and counts requires $\alpha_{\rm CE}$ well above unity, even more so if other formation channels contribute. The strong dependence on the delay-time distribution suggests that revising the mass-transfer prescriptions, and CE evolution in particular, could substantially change this picture.

\section{Summary}
\label{sec:summary}

In this work, we asked how much information binary black hole (BBH) number counts alone can provide in the Einstein Telescope era, within a fully forward-modeled framework. We combined a data-driven, metallicity-resolved cosmic star formation history with {\footnotesize\textup{\textrm{SEVN}}} population-synthesis simulations of isolated binaries, for six values of the common-envelope efficiency $\alpha_{\rm CE}\in[0.5,5]$, two fundamental metallicity relations (AM13 and Cu23) and three treatments of the delay-time distribution (F0, F1, F2). Our main results are the following.

\begin{enumerate}

\item In our fiducial model (AM13 F0, $H_0=70$\,km\,s$^{-1}$\,Mpc$^{-1}$, $\alpha_{\rm CE}=3$) the 10\,km triangular ET detects $N_{\rm det}\simeq10^{4.99}$ BBHs per year, about 60\% of all mergers.

\item Counts alone do not constrain $H_0$, but they restrict $(H_0,\alpha_{\rm CE})$ to a narrow degeneracy curve. If the model is exact, we recover $\alpha_{\rm CE}=3.00^{+0.13}_{-0.17}$ for a flat prior $H_0\in[60,80]$\,km\,s$^{-1}$\,Mpc$^{-1}$, and $\alpha_{\rm CE}=3.006^{+0.027}_{-0.024}$ with a 1\% prior on $H_0$ (both at $2\sigma$). Instead, for a fiducial $\alpha_{\rm CE}=1$, the error on $\alpha_{\rm CE}$ is about 40\% with a flat prior on $H_0$ and about 7\% with the 1\% prior. These values are the statistical floor of number counts, reached only if all other ingredients are known, since any ingredient that changes the merger-rate normalization is degenerate with $\alpha_{\rm CE}$.

\item This precision is sensitive to variations in the population model. If the mock data come from AM13 F0, analyzing them with a different FMR or delay-time treatment shifts the posterior by about 25 (Cu23 F0, F1) to 100 (F2) Poisson widths. Calibration tests on many mock datasets flag every model but the true one as miscalibrated. F2 overestimates $\alpha_{\rm CE}$ (bias $\approx$ $-1.6$ and $-2.0$ for AM13 and Cu23), while Cu23 F0 and F1 underestimate it ($0.55$ and $0.37$). AM13 F0 and F1 are instead very similar: their biases are small ($-0.04$ and $-0.08$) and their posteriors overlap for a fiducial $\alpha_{\rm CE}=3$. This overlap disappears for other fiducial values, e.g. $\alpha_{\rm CE}=1$.

\item Our models overpredict the GWTC-5.0 local merger rate, $\mathcal{R}^{\rm LVK}_{0.2}=30.4^{+6.5}_{-5.4}$\,Gpc$^{-3}$\,yr$^{-1}$, by a factor of 4 to 9 (F0), or 1.5 to 7 for a log-flat delay-time distribution (F2). The excess persists if we replace our SFR density with the Madau \& Fragos fit and a metallicity spread consistent with galaxy observations, so it cannot be attributed to the star formation history alone. It comes mainly from Channel~I, i.e. binaries that undergo common-envelope evolution after stable mass transfer. Matching GWTC-5.0 would require $\alpha_{\rm CE}\approx13$ (F0) or $\approx6$ (F2), which points towards a revision of the CE and mass-transfer prescriptions in population synthesis simulations and not necessarily towards extreme parameter values.
\end{enumerate}

A number-count analysis is only a first step toward a full spectral-siren inference, which we will present in a follow-up paper. It already offers a fast and cheap test of forward models. With $\sim10^5$ events per year, ET will be sensitive to any modeling ingredient that changes the predicted number of detections by more than about 0.3\%.

\acknowledgments

MS and GA acknowledge financial support from the Istituto Nazionale di Fisica Nucleare (INFN), Sezione di Trieste, through the ET-Italia and TEONGRAV initiatives. MS acknowledges financial support from the Istituto Nazionale di Fisica Nucleare (INFN), Sezione di Trieste, through the VIRGO initiative. MS acknowledges support from the INAF-Large Grant 2024: “Envisioning Tomorrow: prospects and challenges for multimessenger astronomy in the era of Rubin and Einstein Telescope”. AL has been supported by the Istituto Nazionale di Fisica Nucleare (INFN) via the specific national initiative QGSKY. MB acknowledges that this article was produced while attending the PhD program in PhD in Space Science and Technology at the University of Trento, Cycle XXXIX, with the support of a scholarship financed by the Ministerial Decree no. 118 of 2nd March 2023, based on the NRRP - funded by the European Union - NextGenerationEU - Mission 4 "Education and Research", Component 1 "Enhancement of the offer of educational services: from nurseries to universities” - Investment 4.1 “Extension of the number of research doctorates and innovative doctorates for public administration and cultural heritage” - CUP E66E23000110001 and support by the Italian grant Project SPACE-IT-UP by the Italian Space Agency and Ministry of University and Research, Contract Number 2024-5-E.0. S.M and A.S. are supported by ERC grant GravitySirens  101163912. Funded by the European Union. Views and opinions expressed are however those of the author(s) only and do not necessarily reflect those of the European Union or the European Research Council Executive Agency. Neither the European Union nor the granting authority can be held responsible for them. This material is based upon work supported by NSF's LIGO Laboratory which is a major facility fully funded by the National Science Foundation.

\appendix
\section{Number likelihood as a marginalized hierarchical likelihood}\label{app:marg_like}
Given $N_{\rm obs}$ observed events $x_i$ in a catalog, the hierarchical likelihood is found to be \cite{Vitale:2022hier_like1, Mandel:2019hier_like2}:
\begin{equation}
\mathcal{L}(x_1,...,x_{N_{\rm obs}}, N_{\rm obs}|\Lambda)
\propto
e^{-N_{\rm det}(\Lambda)}
N_{\rm det}(\Lambda)^{N_{\rm obs}}
\prod_{i=1}^{N_{\rm obs}}
\frac{
\int \mathrm{d}\theta\,
\mathcal{L}_{\rm GW}(x_i|\theta,\Lambda)
\pi_{\rm pop}(\theta|\Lambda)
}{
\int \mathrm{d}\theta\,
p_{\rm det}(\theta,\Lambda)
\pi_{\rm pop}(\theta|\Lambda)
}.
\label{eq:hier_like_app}
\end{equation}
with $\Lambda$ the hyper-parameters of the population model $\pi_{\rm pop}(\theta|\Lambda)$, $\mathcal{L}_{\rm GW}$ the single-event gravitational wave likelihood, $p_{\rm det}(\theta,\Lambda)$ the detection function (Eq.~\ref{eq:pdet}) and $N_{\rm det}$ the number of detectable events (Eq.~\ref{eq:N_det}).

By marginalizing Eq.~\ref{eq:hier_like_app} at fixed $N_{\rm obs}$ over every single event within the space of detectable events we obtain:
\begin{equation*}
    \int_{x_1,...,x_{N_{\rm obs}}\in \text{detectable}} \text{d}x_1\cdots\text{d}x_{N_{\rm obs}} \mathcal{L}(x_1,...,x_{N_{\rm obs}}, N_{\rm obs}|\Lambda) \propto
\end{equation*}
\begin{equation*}
    \propto e^{-N_{\rm det}(\Lambda)}
N_{\rm det}(\Lambda)^{N_{\rm obs}} 
\int_{x_1,...,x_{N_{\rm obs}}\in \text{detectable}} \text{d}x_1\cdots\text{d}x_{N_{\rm obs}} \prod_{i=1}^{N_{\rm obs}}
\frac{
\int \mathrm{d}\theta\,
\mathcal{L}_{\rm GW}(x_i|\theta,\Lambda)
\pi_{\rm pop}(\theta|\Lambda)
}{
\int \mathrm{d}\theta\,
p_{\rm det}(\theta,\Lambda)
\pi_{\rm pop}(\theta|\Lambda)
} \propto 
\end{equation*}
\begin{equation*}
\propto e^{-N_{\rm det}(\Lambda)}
N_{\rm det}(\Lambda)^{N_{\rm obs}} 
\prod_{i=1}^{N_{\rm obs}}
\int_{x_i\in \text{detectable}} \text{d}x_i 
\left[\frac{
\int \mathrm{d}\theta\,
\mathcal{L}_{\rm GW}(x_i|\theta,\Lambda)
\pi_{\rm pop}(\theta|\Lambda)
}{
\int \mathrm{d}\theta\,
p_{\rm det}(\theta,\Lambda)
\pi_{\rm pop}(\theta|\Lambda)
}\right] \propto 
\end{equation*}
\begin{equation}
    \propto e^{-N_{\rm det}(\Lambda)} N_{\rm det}(\Lambda)^{N_{\rm obs}} = \mathcal{L}(N_{\rm obs}|\Lambda)\,,
    \label{eq:numberlike}
\end{equation}
where we used Eq.~\ref{eq:pdet} $\forall x_i$.

\section{Finding the correct number of injections}\label{app:N_gen}
In subsection \ref{sec:BBH_NC} we stated the criterion $N_{\rm eff,inj} > 4 N_{\rm obs}$, from Ref. \cite{Farr:2019Neff_inj}, on the number of effective injections $N_{\rm eff,inj}$ defined by Eq.~(\ref{eq:N_eff_inj}) and using the number of observed events $N_{\rm obs}$. In the following we will prove that choosing the number of generated injections as $N_{\rm gen}=4N_{\rm id}$, where $N_{\rm id}$ is the total number of events for an ideal detector observed in a time $T_{\rm obs}$, is enough to respect the criterion.

When we generate mock catalogs of observed events in subsection \ref{sec:BBH_NC} we first draw $N_{\rm obs}$ observed events from a Poisson process with mean $N_{\rm det}$, i.e. the number of detected events in a time $T_{\rm obs}$ from Eq.~(\ref{eq:N_det}); thus $N_{\rm obs} \sim \mathcal{P}oisson(N_{\rm det})$ \cite{Vitale:2022hier_like1}.

Defining the ratio $r = N_{\rm thr}/N_{\rm gen}$ we have $N_{\rm det} = N_{\rm id} r$ from Eq.~(\ref{eq:approx_N_det2}) (with $N_{\rm thr}$ the number of injections passing the SNR threshold out of $N_{\rm gen}$ generated ones) and if $r$ remains approximately constant once a sufficient number of injections has been generated, we obtain the estimate 
\begin{equation}
    N_{\rm obs}\approx N_{\rm id}\,r \pm n\sqrt{N_{\rm id}\,r}\,,
\end{equation}
where the second term represents the possible negative or positive fluctuations of $n$ standard deviations for a Poisson variable around its mean.
The requirement $N_{\rm eff,inj} > 4 N_{\rm obs}$ then leads to

\begin{equation}
    N_{\rm gen} \dfrac{r}{1-r} > 4N_{\rm id} r \pm 4 n\sqrt{N_{\rm id}r}\,.
    \label{eq:ineq}
\end{equation}

A convenient choice satisfying this condition is $N_{\rm gen}=4N_{\rm id}$. Substituting this value into the inequality yields
\begin{equation}
    N_{\rm id} > n^2 \dfrac{(1-r)^2}{r^3}\,.
    \label{eq:check}
\end{equation}

\begin{figure}[h]
    \centering
    \includegraphics[width=1\textwidth]{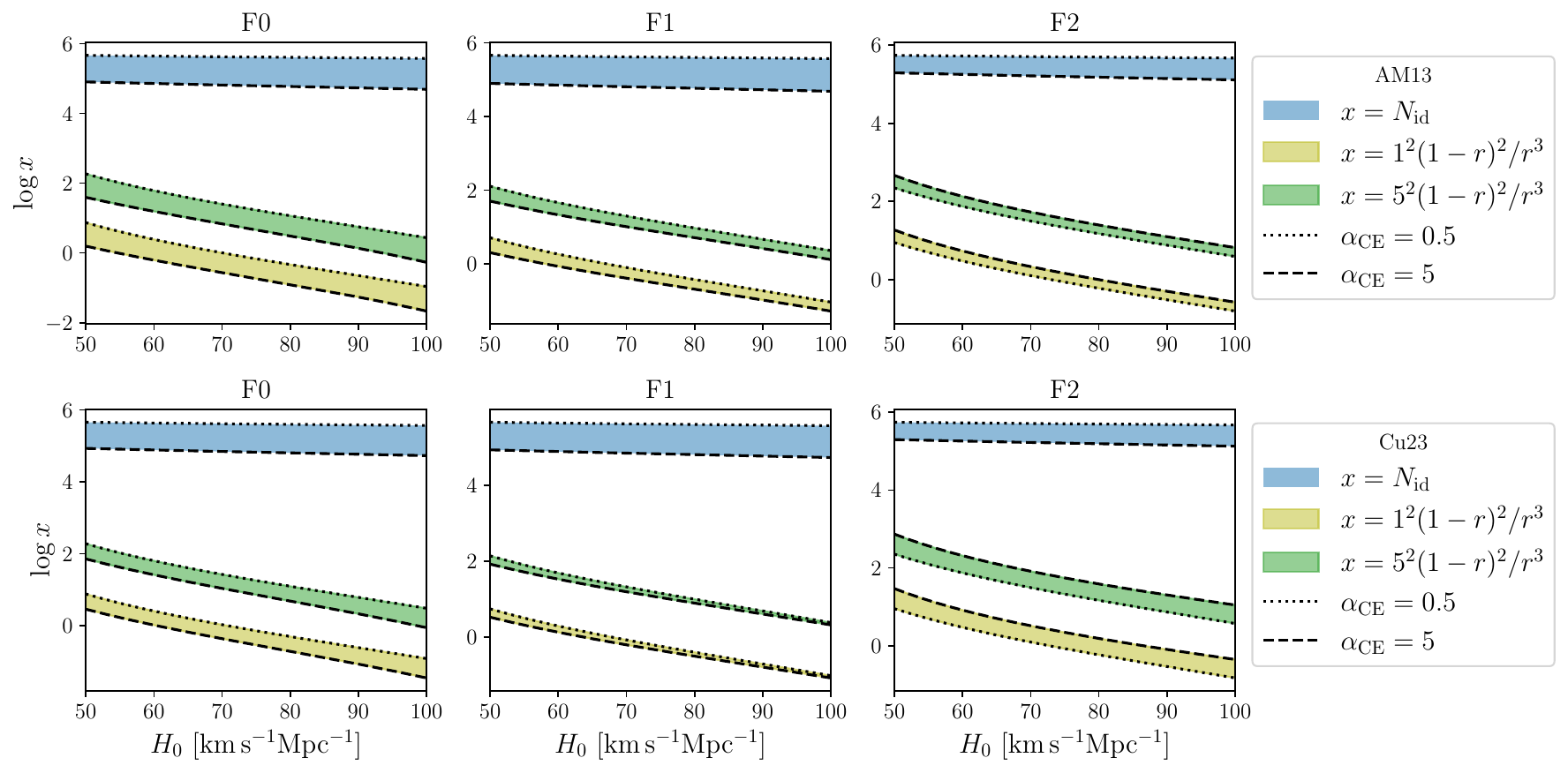}
    \caption{Comparison between $N_{\rm id}$ and the threshold $n^2(1-r)^2/r^3$ from Eq.~(\ref{eq:check}) as a function of $H_0$ and a range of $\alpha_{\rm CE}$ from 0.5 (dotted lines) to 5 (dashed lines). The blue bands show the predicted number of mergers for an ideal detector, $N_{\rm id}$, while the yellow and green bands represent the right-hand side of Eq.~(\ref{eq:check}), $n^2(1-r)^2/r^3$, respectively for the extreme values $n=1$ and $n=5$. Columns correspond to the three merger density function factorizations (F0, F1, F2), while rows correspond to the two FMR prescriptions (AM13 and Cu23). }
    \label{fig:AM13_crit} 
\end{figure}

To verify that this condition is satisfied, we compare the two sides of the inequality for all our merger-rate models in Fig.~\ref{fig:AM13_crit}. The figure shows that the predicted values of $N_{\rm id}$ (blue bands) are well above the right-hand side of Eq.~(\ref{eq:check}) (yellow to green bands as $n=1-5$) across the explored range of $H_0$ and spanning the range $\alpha_{\rm CE}\in[0.5, 5]$ in the bands, confirming that the choice $N_{\rm gen}=4N_{\rm id}$ provides a sufficient number of injections for our analysis.

\section{Statistical miscalibration tests}\label{sec:pp_plots}
We identify model miscalibration by means of a statistical test on the expected distribution of credible regions containing at most the true values. We consider the prior $\pi(\Lambda)$ and posterior $P(\Lambda)$ on the hyper-parameters and use them as follows:
\begin{itemize}
    \item[1)] Construct $K$ mock datasets by drawing $\Lambda_k\sim \pi(\Lambda)$ from the prior, with $k=1,...,K$ and extract $K$ mock observations (1000 used for the analysis) $N_{\text{obs},k}^{M_0}\sim\mathcal{P}oisson(N_{\rm det}^{M_0}(\Lambda_k))$ with a fixed FMR-factorization model ($M_0$=AM13 F0 in our case).
    \item[2)] Compute the posterior distributions using each observation at different FMR-factorization model $M$:
    
    \begin{equation}
    P^M_k(\Lambda) = P(\Lambda|N_{\text{obs},k}^{M_0},N_{\rm det}^M) \propto N_{\rm det}^M(\Lambda)^{N_{\text{obs},k}^{M_0}}\,e^{-N_{\rm det}^M(\Lambda)}\pi(\Lambda)\,.
    \end{equation}
    \item[3)] Define the random variables (RVs) $L_k^M = P_k^{M}(\Lambda)$, their cumulative distribution $F_{L_k^{M}}(t) = \text{Pr}(L_k^{M} < t)$ and the set of $K$ outcomes from the RV with $M=M_0$, i.e. $\ell_k=P_k^{M_0}(\Lambda_k)$. We find that for a number $f\in[0,1]$ 
    \begin{equation}
    \text{Pr}(F_{L_k^{M}}(L_k^{M_0})< f) = \text{Pr}(L_k^{M_0} < F_{L_k^{M}}^{-1}(f)) = F_{L_k^{M_0}}(F_{L_k^{M}}^{-1}(f)) \underset{\text{iff } M=M_0}{=}  f\,;
    \end{equation}
    therefore $F_{L_k^{M}}\sim \mathcal{U}(0,1)$ only when the model is calibrated $(M=M_0)$.
    \item[4)] Find the credible levels $(\beta_k^M)$ of the highest density posterior regions that contain the truth $\Lambda_k$ at its border:
    \begin{equation}
    \beta_k^M = \int_{\{\Lambda| P_k^M(\Lambda)\geq P_k^M(\Lambda_k)\}} P^M_k(\Lambda) \text{d}\Lambda\,;
    \end{equation}
    which are directly linked to the joint posteriors, and we see that when $M=M_0$
    \begin{equation}
    \beta_k^{M_0} = \text{Pr}_{\Lambda\sim P_k^{M_0}} (L_k^{M_0}\geq \ell_k) = 1 - F_{L_k^{M_0}}(\ell_k)\,\Rightarrow \beta_k^{M_0}\sim \mathcal{U}(0,1)\,;
    \end{equation}
    \item[5)] We identify miscalibration with a Kolmogorov-Smirnov test on the $\beta^M$ distribution. 
\end{itemize}

\bibliographystyle{JHEP}
\bibliography{biblio}

\end{document}